\documentclass{aa}
\usepackage[varg]{txfonts}
\usepackage{graphicx}
\usepackage[switch]{lineno}
\usepackage{amsmath,wasysym}

\bibpunct{(}{)}{;}{a}{}{,}

\begin{document}

\title{Formation of starspots in self-consistent global dynamo models: Polar spots on cool stars}

\author{Rakesh K. Yadav\inst{\ref{MPS}}$^{,}$\inst{\ref{IAG}} 
\thanks{\emph{Email: yadav@mps.mpg.de}}
\and Thomas Gastine\inst{\ref{MPS}}
\and Ulrich R. Christensen\inst{\ref{MPS}}
\and Ansgar Reiners\inst{\ref{IAG}}
}

\institute{Max-Planck-Institut f\"{u}r Sonnensystemforschung, Justus-von-Liebig-Weg 3, 37077 G\"{o}ttingen, Germany\label{MPS}
\and
Institut f\"ur Astrophysik, Georg-August-Universit\"at, Friedrich-Hund-Platz 1, 37077 G\"ottingen, Germany\label{IAG}
}

\abstract
{Observations of cool stars reveal dark spot-like features on their surfaces. Compared to sunspots, starspots can be bigger or cover a larger fraction of the stellar surface. While sunspots appear only at low latitudes, starspots are also found in polar regions, in particular on rapidly rotating stars. Conventional flux-tube models have been invoked to explain starspot properties. However, these models use several simplifications and so far the generation of either sunspots or starspots has not been demonstrated in a self-consistent simulation of stellar magnetic convection.}
{To clarify conditions for the spontaneous formation of dark spots in numerical models of convection-driven stellar dynamos.}
{We simulate convection and magnetic field generation in rapidly rotating spherical shells under the anelastic approximation. The high-resolution simulations were performed using a fully-spectral magnetohydrodynamic code.}
{We demonstrate for the first time that a self-consistent distributed dynamo can spontaneously generate high latitude dark spots. Dark spots are generated when large-scale magnetic field, generated in the bulk of the convection zone, interacts with and locally quenches flow near the surface. Prerequisites for the formation of sizeable dark spots in the model are sufficiently strong density stratification and rapid rotation. }
{Our models present an alternative scenario for starspot formation by distributed dynamo action. Our results also lend strong support to the idea that dynamos in the interiors of rapidly rotating stars might be fundamentally different from the solar one.}


\titlerunning{Starspot formation}
\authorrunning{Yadav et al.}

\maketitle

\section{Introduction}
Dark spots are one of the most remarkable features on the surface of the Sun. Sunspots are believed to be caused by the interaction of solar magnetic field with near surface turbulent convection: in regions of strong magnetic field convection is highly quenched which leads to inefficient transport of heat, forming local cool and dark spots \citep[see][]{stein2012}. The size and distribution of regions with strong magnetic field are ultimately governed by the  underlying dynamo mechanism. In the current popular picture of the solar dynamo the interface region of strong radial shear between the radiative core and the convective envelope, the tachocline, is thought to be instrumental~\citep{charbonneau2005}. The tachocline creates strong toroidal magnetic field by shearing the poloidal field lines. Thin flux-tubes of strong magnetic field detach from the tachocline and rise to the solar surface where they provide the necessary magnetic field for sunspots. 

In the past few decades, observational techniques, e.g. photometric light-curve modelling and Doppler imaging~\citep{vogt1987}, have been applied to infer dark starspots on other cool stars. Light-curve modelling is inherently ambiguous\footnote{However, see \citet{davenport2014} who use light curves featuring starspot-occulting planetary transits to disentangle some of the degeneracies of light curve modelling.} and gathering good quality data for Doppler imaging is rather tedious. The evolution of starspots (appearing and disappearing during the observations) makes it even more complicated. Nonetheless, there are a few robust features which have gathered support over the years (see \citet{berdyugina2005}, \citet{strassmeier2009}, and reference therein). A particularly intriguing feature is the occurrence of starspots near the rotational poles of many rapidly rotating stars~\citep{strassmeier2002}. This behaviour is in stark contrast with sunspots which appear exclusively at low latitudes. Starspots in these stars also appear to be rather large,  sometimes covering as much as a few percent of the stellar surface, while the collective area of sunspots even during the solar maximum covers only a small fraction of a percent of the solar photosphere.

Extending the flux-tube models to rapidly rotating cool stars, it has been suggested that rising flux-tubes appear at high latitudes either due to the influence of strong Coriolis forces~\citep{schussler1992,isik2011} or that they appear at low latitudes and are then advected polewards by near-surface flows~\citep{schrijver2001, holzwarth2006, isik2007, isik2011}. Mechanism based on magnetic Rossby waves in the tachocline of rapidly rotating stars have also been proposed for polar spot formation~\citep{zaqarashvili2011}.

In contemporary flux-tube models the {\em formation} of dark spots is not considered. These models  provide information only about the background magnetic field which  supposedly leads to dark spot formation. On the other hand, direct numerical simulations of distributed dynamos and dark spot formation have been rather disconnected. Global dynamo models \citep[e.g.][]{miesch2005, steffen2007, ghizaru2010, kapyla2012, gastine2012b, nelson2013, hotta2014, fan2014} simulate the generation of large scale magnetic fields, while local simulations \citep[e.g.][]{vogler2005, stein2006, rempel2009, kitiashvili2010, cheung2010, stein2011} study the formation of dark spots in the presence of a {\em prescribed} magnetic field. \citet{mitra2014} recently simulated a self-consistent dynamo using an artificially forced flow in a box-geometry; the resulting magnetic field was able to quench flow in a localized region in the box, mimicking dark spot formation. Although this is certainly a step in the right direction, the setup is simplistic and does not incorporate rotation or convection.

The idea that stellar dynamos necessarily rely on strong shear flow in a tachocline region is certainly worth revisiting, given that flux-tube models are contested even in the solar context~\citep{brandenburg2005}. Furthermore, recent high-resolution numerical simulations of distributed dynamos in spherical shells are also providing a rather different perspective on the solar dynamo~\citep{ghizaru2010, kapyla2012, nelson2013}. It has been suggested that dynamos in stars where rotation plays a dominant role might be fundamentally different from the solar case~\citep{christensen2009, donati2009, reiners2012}. Similar to the dynamos thought to be working in planetary interiors, a distributed dynamo in stars that pervades the entire convection zone can potentially avoid many of the shortcomings of the flux-tube models. For instance, it is hard to imagine how a tachocline region around a geometrically small radiative core in stars (e.g. young pre-main sequence stars or low-mass stars) could govern the dynamo. Flux-tube models which have been extended to such stars produce magnetic field {\em only} near the rotational poles~\citep{holzwarth2004}. For fully-convective stars the flux-tube picture would clearly not apply. A distributed dynamo on the other hand does not need a tachocline and hence can easily operate in stars with small or no radiative cores.

In order to investigate the generation of a global magnetic field in spherical shell convection and the simultaneous formation of cool surface spots self-consistently, we further advance a dynamo model that has been applied to gas planets and cool stars~\citep{gastine2012b, gastine2013a}. We do not attempt to model a specific type of star or to match stellar structure and properties as faithfully as possible. Hence this study is in essence an exploratory one to find out which basic ingredients are necessary to spontaneously generate  dark spots in global numerical simulations without a tachocline. In this study our main focus will be to produce large starspots at high latitudes in the framework of distributed dynamo models.

For the case of polar spots it is rather tempting to envisage a scenario where largely axisymmetric and dipolar magnetic fields, similar in geometry to the Earth's or Jupiter's field, are the backbone. In such dynamos the magnetic field strength naturally peaks at high latitudes. We adopt this as our working hypothesis and pursue it for generating polar spots.

\section{Model setup}
\subsection{\label{ane_eq}Anelastic equations}
We use the anelastic magnetohydrodynamic (MHD) equations~\citep{braginsky1995, lantz1999} to simulate the subsonic flows below the photosphere of a star. An electrically conducting fluid convects under the influence of a fixed entropy contrast across a spherical shell with an inner radius $r_{i}$ and outer radius $r_{o}$. The shell rotates with a constant angular velocity $\Omega$ about the $z$-axis. 

The model equations are non-dimensionalised using the shell thickness $d= r_{o} - r_{i}$ as the reference length scale and the viscous diffusion time $d^2/\nu$ ($\nu$ is viscosity) as the time scale. The density at the outer boundary $\rho_{o}$ is used as the density unit and $\sqrt{\Omega\lambda\mu\rho_{o}}$ is the magnetic field scale. The pressure is scaled by $\rho_o\nu\Omega$. The magnetic diffusivity $\lambda$ and the magnetic permeability $\mu$ are assumed to be constant. The imposed entropy contrast $\Delta s$ between the inner and the outer boundary defines the entropy scale. 

In the anelastic approximation the thermodynamic variables, density, and pressure, are decomposed into a sum of reference state values and small perturbations as $x(r,\theta, \phi,t) = \tilde{x}(r) + x'(r,\theta, \phi,t)$. The reference state $\tilde{x}(r)$ corresponds to an adiabatic ideal gas. The reference state density $\tilde{\rho}$ and temperature $\tilde{T}$ are then related by $\tilde{\rho}=\tilde{T}^{m}$ where $m$ is polytropic index. Assuming a linear variation of gravity the reference state temperature is given by
\begin{gather}
\tilde{T} = 1 -  \frac{Di}{2(1-\eta)} \left( \frac{r^{2}}{r_{o}^{2}}-1 \right), 
\end{gather}
where
\begin{gather}
Di = \frac{g_o\,d}{c_p\,T_o} =2\, \frac{e^{\frac{N_{\rho}}{m}} - 1}{1+\eta} \nonumber
\end{gather} 
is the dissipation number, $g_{o}$ is the gravity at the outer boundary, $c_{p}$ is the specific heat at constant pressure, $N_{\rho}=\ln(\tilde{\rho}(r_{i})/\tilde{\rho}(r_{o}))$ is the number of density scale heights across the shell, and $\eta=r_{i}/r_{o}$ is the aspect ratio of the spherical shell.

The non-dimensional evolution equation for velocity ${\bf u}$ is:
\begin{gather}
E\left(\frac{\partial{\bf u}}{\partial t}+{\bf u}\cdot\nabla\mathbf{u}\right)= -\nabla\frac{p'}{\tilde{\rho}}-2\,\hat{z}\times{\bf u}+\frac{Ra\, E}{Pr}\, \frac{r}{r_{o}}\, s\,\hat{r}\,+ \nonumber \\ 
\,\frac{1}{Pm\,\tilde{\rho}}(\nabla\times{\bf B})\times{\bf B}+\frac{E}{\tilde{\rho}}\nabla\cdot \mathsf{S}, \label{eq:vel} 
\end{gather}
where $p$ is the pressure, $\mathbf{B}$ is the magnetic field, $s$ is the specific entropy, and $\hat{r}$ is the radial unit vector.  
\begin{gather}
\mathsf{S}_{ij}=\tilde{\rho}\left(\frac{\partial u_{i}}{\partial x_{j}}+\frac{\partial u_{j}}{\partial x_{i}}-\frac{2}{3}\delta_{ij}\nabla\cdot\mathbf{u}\right) 
\end{gather}
is the traceless rate-of-strain tensor.
The entropy is governed by
\begin{gather}
\tilde{\rho}\tilde{T}\left(\frac{\partial s}{\partial t}+{\bf u}\cdot\nabla s\right)=\frac{1}{Pr}\nabla\cdot(\tilde{\rho}\tilde{T}\nabla s)\,+ \frac{Pr\,Di}{Ra}\left[\mathsf{S}^2 + \frac{1}{E\,Pm^2}(\nabla\times\mathbf{B})^{2}\right]. \label{eq:entropy} 
\end{gather}
The magnetic field ${\bf B}$ evolves according to the induction equation
\begin{gather}
\frac{\partial{\bf B}}{\partial t}=\nabla\times({\bf u}\times{\bf B})+\frac{1}{Pm}\nabla^{2}{\bf B}. \label{eq:mag} 
\end{gather}
The set of equations is completed by
\begin{gather}
\nabla\cdot {\bf B} = 0\,\, \textrm{and}\,\, \nabla\cdot(\tilde{\rho} {\bf u}) = 0.
\end{gather}

The non-dimensional control parameters appearing in above equations are: 
\begin{gather}
\text{the magnetic Prandtl number }Pm=\frac{\nu}{\lambda}, \nonumber \\
\text{the Prandtl number } Pr=\frac{\nu}{\kappa}, \nonumber \\
\text{the Rayleigh number } Ra=\frac{g_{o}\,d^{3}\,\Delta s}{c_{p}\,\nu\,\kappa}, \nonumber \\ 
\text{the Ekman number } E=\frac{\nu}{\Omega\,d^{2}}, \nonumber
\end{gather}
where $\kappa$ is thermal diffusivity. The transport coefficients $\nu$, $\lambda$, $\kappa$ are assumed constant. To better model magnetic field driven structures in outer layers we have used a constant $\lambda$. In contrast, earlier studies of stellar dynamos usually assumed a radially increasing $\lambda$, making coupling of flow and magnetic field weaker in outer layers \citep[e.g.][]{browning2008, fan2014}. The aspect ratio $\eta=r_{i}/r_{o}$ is set to 0.35 for all the simulations reported below.

Following earlier studies \citep[e.g.][]{fan2014} we also report the various energy fluxes which appear in systems governed by the anelastic equations: the entropy diffusion flux $F_{diff}$, the enthalpy flux $F_{conv}$, the kinetic energy flux $F_{KE}$, the viscous diffusion flux $F_{visc}$, the Poynting flux $F_{poyn}$, and the resistive flux $F_{res}$. In our non-dimensional units the various fluxes are defined as follows:
\begin{gather}
F_{diff}=-\frac{1}{Pr}\tilde{\rho}\tilde{T}\left[\nabla s\right]_{r},  \label{eq:flux1} \\
 F_{conv}=\tilde{\rho}\tilde{T}s\,u_{r}+\frac{Pr\,Di}{E\,Ra}\, P'u_{r},  \\
 F_{KE}=\frac{Pr\,Di}{Ra}\, u_{r}\left(\frac{\tilde{\rho}u^{2}}{2}\right),  \\ 
 F_{visc}=-\frac{Pr\,Di}{Ra}\,\left[\mathbf{u}\cdot \mathsf{S}\right]_{r},  \\
 F_{poyn}= -\frac{Pr\,Di}{Ra\,E\,Pm}\left[(\mathbf{u}\times\mathbf{B})\times\mathbf{B}\right]_{r},  \\
 F_{res}= \frac{Pr\,Di}{Ra\,E\,Pm^2}\left[(\nabla\times\mathbf{B})\times\mathbf{B}\right]_{r},  \label{eq:flux6} 
\end{gather}
where $[...]_r$ represents the radial component.

\subsection{Boundary conditions}
Both inner and outer boundaries are impenetrable, stress-free, and electrically insulating. Constant entropy is assumed on both boundaries. Unlike previous numerical studies of stellar convection~\citep{browning2006, ghizaru2010, kapyla2010, masada2013} we do not model a convectively stable layer below the convection zone. The region below the inner boundary is treated as static.

It is worth noting that our choice of constant entropy on the outer boundary allows the possibility of forming dark spots where heat flux is lower than neighbouring regions. This was not possible in earlier studies of stellar convection zones where constant heat flux boundary conditions were employed.

\subsection{Initial conditions}
Global dynamo simulations~\citep{simitev2009, sasaki2011, schrinner2012, gastine2013a, gastine2012b, yadav2013a} in spherical shells with free-slip boundaries show bistability. In the bistable regime dynamos started with a weak and small-scaled magnetic field saturate at a multipolar and non-axisymmetric magnetic field, while the ones started with a magnetic field having a strong axial dipole saturate in a dipole-dominant field configuration. We use either of these magnetic field configurations as initial condition to explore different dynamo solutions for a given set of control parameters.

\subsection{Numerical technique}
The model equations are solved using the MHD-code MagIC~\citep{wicht2002, gastine2012a} which has been tested using community-based benchmark simulations~\citep{jones2011an}. After decomposing the mass flux and magnetic field into toroidal and poloidal components as 
\begin{gather}
\tilde{\rho}\mathbf{u}=\nabla\times(\nabla\times W\hat{r}) + \nabla\times X\hat{r}, \nonumber \\ 
\mathbf{B}=\nabla\times(\nabla\times Y\hat{r}) + \nabla\times Z\hat{r}, \nonumber
\end{gather}
the scalar potentials $W,X,Y,Z$, the pressure, and the entropy are then expressed in terms of spherical harmonics in longitude and latitude and Chebyshev polynomials in radius. The system of equations is time-advanced using an explicit second-order Adams-Bashforth scheme for Coriolis and non-linear terms and an implicit Cranck-Nicolson scheme for the rest of the terms~\citep{glatzmaier1984,Christensen2007}.

\subsection{\label{con_param}Control parameters}
Capturing the dynamics of rotationally dominated large-scale convection in stellar interiors is the primary aim of this study and the control parameters we chose reflect this to some extent. However, the technical feasibility severely constrains our control parameter choice. Hence, this study, or any global numerical study for that matter, implicitly assumes that the unresolved small-scale turbulence does not affect the large scale properties of the system. The control parameters for different setups are tabulated in Tab.~\ref{tab:tab1}.

\begin{table*}
\begin{center}
\tiny
\caption{Simulation setups} \label{tab:tab1} 
\begin{tabular}{ccccccccccccccc}
\hline 
\hline
Model & $Pr$ & $Pm$ & $E$ & $N_{\rho}$ & m & $Ra$ & $N_\phi\times N_\theta\times N_r$ & TKE & KE$^{pol}$ & KE$^{tor}$ & KE$^{DR}$ & ME$^{pol}$ & ME$^{tor}$ & $Nu$\tabularnewline
\hline 
H-Pr10-N5 & 10 & - & $3\times10^{-4}$ & 5 & 2 & $3\times10^{7}$ & 1280$\times$640$\times$121 & 5.6$\times10^4$ & 0.38 & 0.62 & 0.18 & - & - & 3.2\tabularnewline

NA-Pr10-N5 & 10 & 10 & $3\times10^{-4}$ & 5  & 2 & $3\times10^{7}$ & 1440$\times$720$\times$121 & 5.2$\times10^4$ & 0.41  & 0.59 & 0.08 &  0.15  & 0.14  & 3.2\tabularnewline
  
A-Pr10-N5 & 10 & 10 & $3\times10^{-4}$ & 5  & 2 & $3\times10^{7}$ & 1440$\times$720$\times$121 & 4.6$\times10^4$ & 0.45  & 0.55  & 0.03 & 1.36  & 1  & 3.2\tabularnewline
 
A-Pr10-N4 & 10 & 10 & $3\times10^{-4}$ & 4  & 2 & $2\times10^{7}$ & 1024$\times$512$\times$121 & 3.7$\times10^4$ & 0.46 &  0.54 & 0.02 & 1.2 & 0.94 & 4.6\tabularnewline
 
A-Pr10-N3 & 10 & 10 & $3\times10^{-4}$ & 3  & 2 & $1.5\times10^{7}$ & 768$\times$384$\times$97 & 3$\times10^4$ & 0.44  & 0.56  & 0.03 & 0.89  & 0.77  & 6.4\tabularnewline

NA-Pr1-N6 & 1 & 2 & $3\times10^{-4}$ & 6  & 1.5 & $4\times10^{7}$ & 1280$\times$640$\times$121 & 2.3$\times10^6$ & 0.4  & 0.6  & 0.18 &  0.26 & 0.29 & 1.4\tabularnewline

S-Pr1-N5 & 1 & 3 & $3\times10^{-3}$ & 5  & 2 & $10^{7}$ & 2048$\times$1024$\times$121$^{*}$ & 5.4$\times10^6$ & 0.43  & 0.57  & 0.45 & 0.03  & 0.03  & 2.1\tabularnewline
\hline 
\end{tabular}
\tablefoot{Each model's name highlights its  main characteristics: the letter `H' stands for hydrodynamic case, and letters `A', `NA', `S' are for axisymmetric, non-axisymmetric, and small-scaled, respectively, highlighting the nature of the dynamo. The `Pr' and `N' in a model's name stand for the $Pr$ and $N_{\rho}$; the corresponding values for these control parameters follows. TKE is the volume-integrated non-dimensional kinetic energy defined as $0.5\int{\rho\,u^2\,dv}$. The KE$^{pol,tor}$ and ME$^{pol,tor}$ are the poloidal and toroidal component of the kinetic and magnetic energy, respectively, normalized by TKE. KE$^{DR}$ is the total kinetic energy (normalized by TKE) contained in the axisymmetric differential rotation. Nusselt number $Nu$ is the ratio of the total heat flux to the conducted heat flux transported from the bottom of the convection zone to the top. Model S-Pr1-N5 was run with two-fold symmetry in the azimuthal direction; the effective grid resolution is therefore 1024$\times$1024$\times$121. }
\end{center}
\end{table*}

There have been some unsuccessful attempts at generating an axial-dipole dominated (ADD) magnetic field in global numerical simulations with density stratified convection zones~\citep{dobler2006, browning2008}. This is in stark contrast with the studies of planetary dynamos (usually ignoring density stratification) where ADD solutions are frequently found~\citep{jones2011}. The planetary dynamo  simulations \citep[e.g.][]{christensen2006} persistently show that as the Ekman number, quantifying the importance of viscous effects as compared to rotational ones, is decreased (i.e. increasing rotational influence) ADD magnetic fields become more stable and can be obtained at low values of $Pm$. For example, ADD fields have been found at $Pm$ as low as 0.06 for $E\approx 10^{-6}$ in geodynamo simulations~\citep{christensen2006}. Reaching such small $E$ is anelastic simulations would be a much more demanding task.

Systematic investigation have revealed that as  the density stratification increases in the convection zone, ADD dynamos gradually become unstable~\citep{gastine2012b, schrinner2014}. \citet{schrinner2014} show that for moderate Ekman numbers used in density stratified simulations a high $Pm$ might be favourable for attaining ADD dynamos.  Decreasing the amplitude of differential rotation (in the form of prograde equatorial jets that are typically found in simulations) might also help to stabilize ADD dynamos~\citep{duarte2013, schrinner2014}. The Prandtl number has been shown to affect the amplitude of the equatorial differential rotation~\citep{christensen2002, simitev2005}. Keeping these results in mind we have used relatively high values of $Pr$ and $Pm$ in our simulations (see Tab.~\ref{tab:tab1}). Moreover, the cases with $Pr$ of 10 generate convection with  moderate Reynolds numbers $Re=v\,d/\nu$ ($v$ is some appropriately averaged  velocity). Consequently, a relatively high value of $Pm$ is required in these simulations to attain high enough magnetic Reynolds number $Rm=Re\,Pm$ so that dynamo action occurs. The Ekman number is set to $3\times10^{-4}$ (except for case S-Pr1-N5). It is equivalent to a Taylor number (=$4/E^2$) of about $4\times10^9$. This value allows us to carry out a small parameter study in a manageable time frame while still having rotationally dominated convection in a medium with strong density stratification.

Simulations with density-stratified convection zones have relatively slower and large-scaled flow in high density regions while the flow is rapid and has smaller scales in regions of low density. The latter demands smaller time-steps and higher grid resolution.  Furthermore, to make sure that a dynamo solution is in an equilibrium state the simulation should run for at least a magnetic diffusion time, implying longer run time at higher $Pm$. Within these constraints the maximum density contrast we could afford in simulations with $Pm$=10 was about 150. The polytropic index also had to be increased to 2 (from a more appropriate value of 1.5 for a monoatomic ideal gas) in these cases to avoid a steep drop in density in the outer layers which would require a higher grid resolution to resolve. For a case with $Pm$=2, however, we could reach a density contrast of about 400 with a polytropic index of 1.5. An exception was made for model S-Pr1-N5 which was run for a third of magnetic diffusion time due to the severe computational requirements.

\subsection{Numerical grid resolution}
All the simulations reported in this study were first run on a grid with 768$\times$384$\times$121 grid points, where the three numbers represent grid resolution in longitude, latitude, and radius, respectively. This grid was sufficient to adequately resolve the flow and magnetic field in most of the interior of the shell, but the relatively vigorous convection near the surface remained under-resolved in most cases. We then stepwise increased the grid resolution and ran the simulation long enough to confirm the results. The resolution reported in Tab.~\ref{tab:tab1} is the maximum resolution used for a particular simulation. 

\begin{figure*}[!htb]
\begin{center}
\includegraphics[scale=0.42]{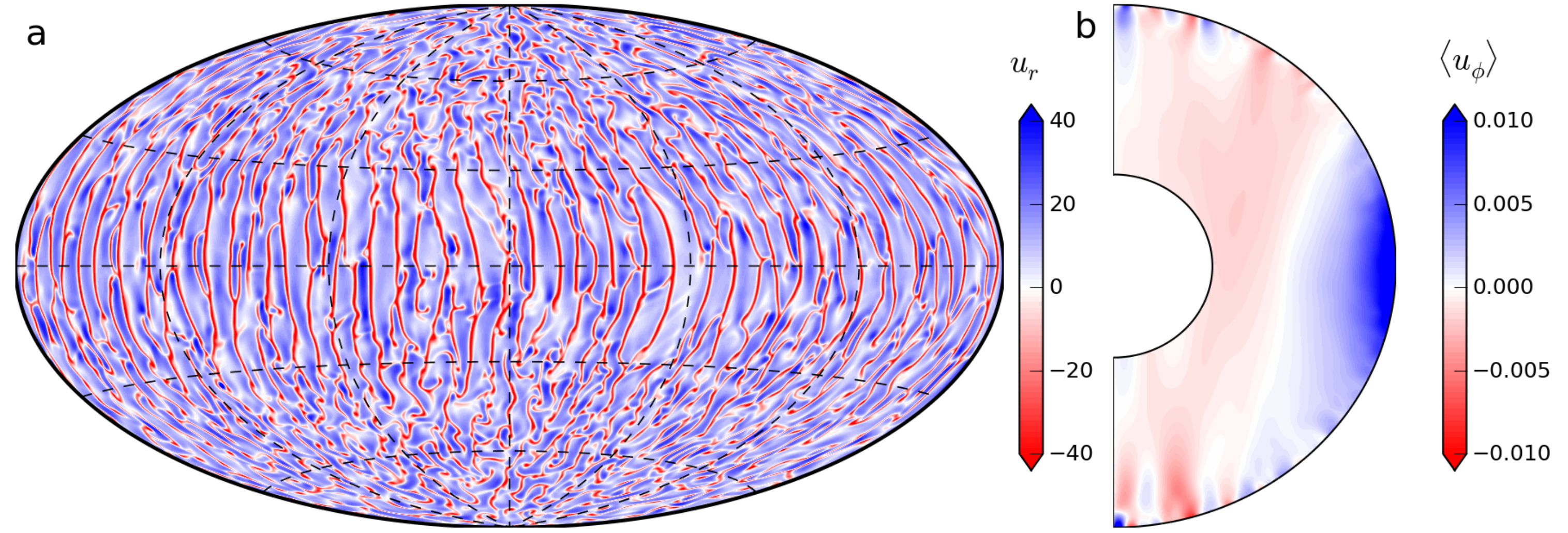} 
\includegraphics[scale=0.14]{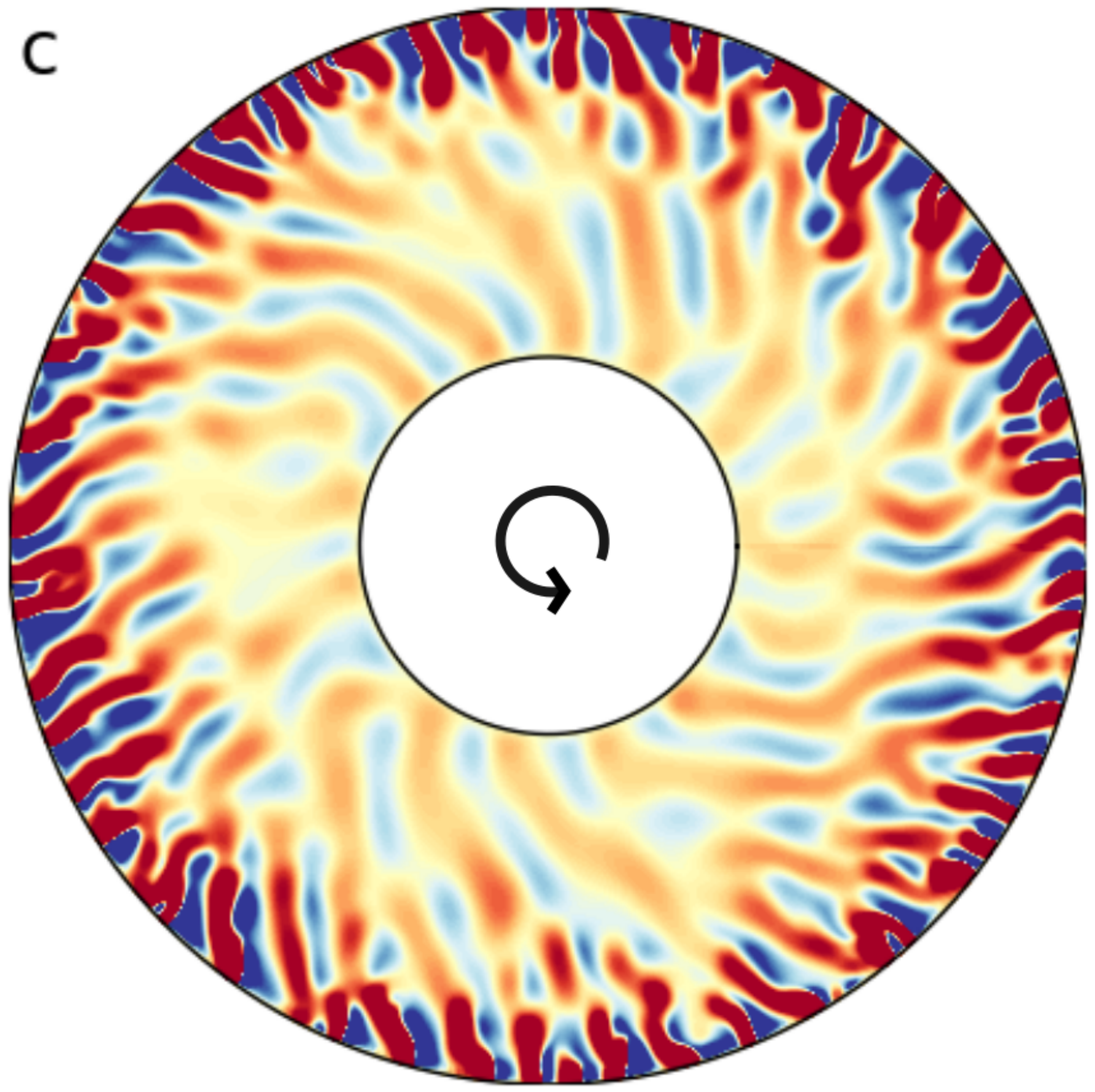}\\
\caption{Flow patterns for hydrodynamical model H-Pr10-N5. Hammer projection of radial velocity $u_{r}$ at $r=0.99\,r_{o}$ is displayed in ({\bf a}) while the longitudinally averaged azimuthal velocity $\langle u_{\phi}\rangle_{\phi}$ (or differential rotation) is given in ({\bf b}). Red and blue indicates downwelling/westward flow and upwelling/eastward flow, respectively. $u_{r}$ and $\langle u_{\phi}\rangle_{\phi}$ are given in terms of the Reynolds number $Re=u\,d/\nu$ and the Rossby number $Ro=u/(d\Omega)$ respectively. The z component of vorticity is shown in ({\bf c}) where red and blue shades represent positive and negative vorticity respectively. The color variations are saturated at values lower than the extrema.}
\label{fig:hydro_flow}
\end{center}
\end{figure*}

\section{Results}
We begin with discussing the results of a simulation for purely hydrodynamical convection and later explore its dynamo action.

\subsection{\label{hydro}Hydrodynamical convection}
To roughly assess how strong rotation influences convection we can use the so-called {\it convective Rossby number} $Ro_{c} = \sqrt{Ra\,E^{2}/Pr}$ \citep[introduced by][]{gilman1977} which estimates the ratio of buoyancy to Coriolis forces. For $Ro_c\ll1$, rotational effects dominate. The model H-Pr10-N5 (see Tab.~\ref{tab:tab1}) has $Ro_{c}\approx0.5$. Therefore convection is likely to be substantially influenced by rotation in this case. Note that  $Ro_{c}$ can be decreased either by decreasing $E$ (computationally very demanding) or increasing $Pr$. The last scenario has been exploited in setting the control parameters for this and most other cases.

Various quantities describing the convection patterns of model H-Pr10-N5  are portrayed in Fig.~\ref{fig:hydro_flow}. The radial velocity $u_{r}$ near the outer boundary shown in ({\bf a}) highlights the typical broad upwellings and narrow downwellings which are formed due to the influence of density stratification. The flow in the deep interior, however, consists of large scale helical columns aligned with the rotation axis, as seen in a 3-D rendering (not shown for this case). Such convection pattern is typical for rotationally dominated convection in simulations of density-stratified convection zones~\citep{miesch2005, browning2008, gastine2012a}. The structural change arises because the influence of rotation varies as a function of radius in a density-stratified fluid and is weaker in the outer layers~\citep{gastine2012a}. In ({\bf b}) the axisymmetric longitudinal velocity $\langle u_{\phi}\rangle_{\phi}$ (angle brackets $\langle...\rangle_{x}$ represent averaging over $x$) or the differential rotation varies only moderately on cylinders aligned with the rotation axis, except near the outer boundary. The typical differential rotation profile, i.e. faster equator and slower poles, is maintained by the Reynolds stresses which arise due to a statistical correlation between radial and longitudinal flow components~\citep{christensen2002, simitev2005}. Reynolds stresses are promoted by the spiralling nature of convection columns which are tilted in the direction of rotation of the shell~\citep{gilman1975,busse1983}. The plot of the axial fluid vorticity $\omega_z=(\nabla\times\mathbf{u})_z$ in the equatorial plane of the shell in ({\bf c}) shows the spiralling columns.

\begin{figure}
\begin{center}
\includegraphics[scale=0.45]{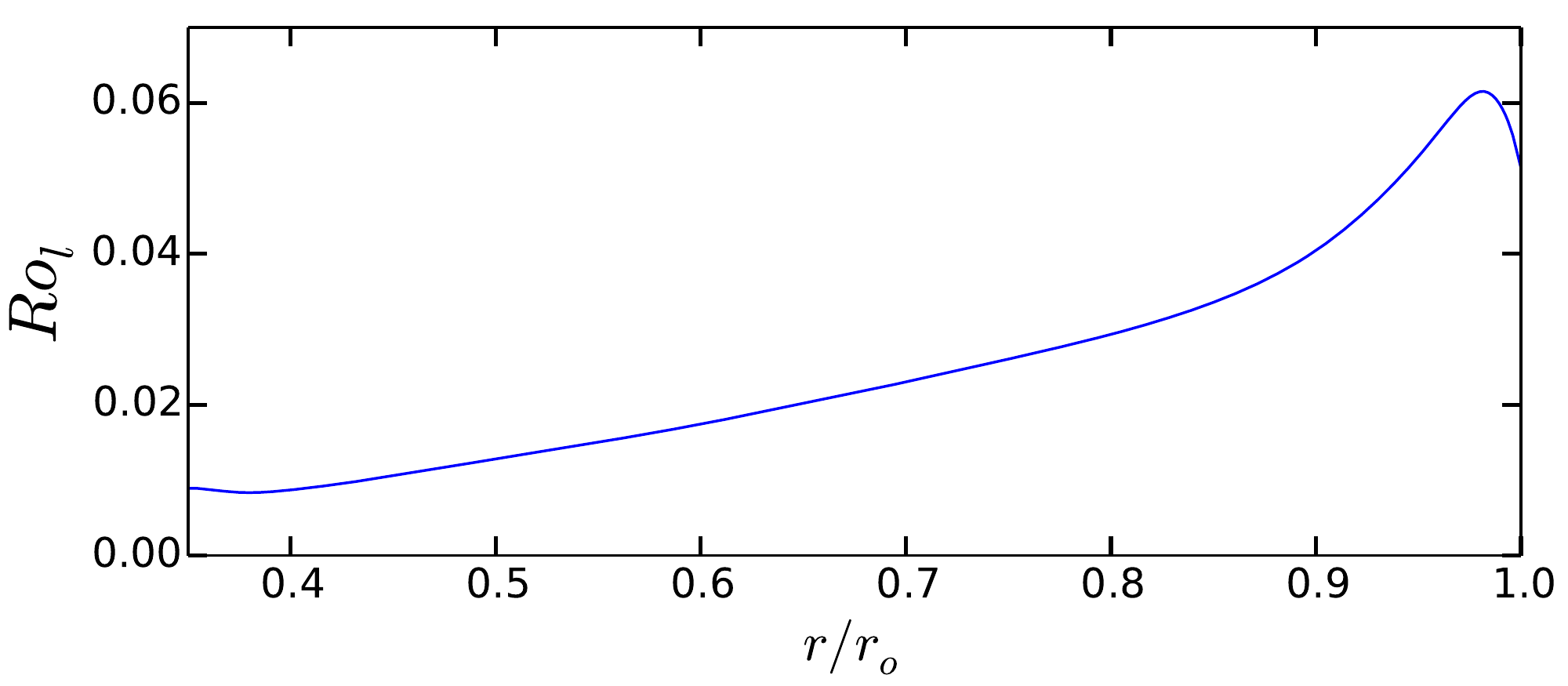}
\caption{Variation of the {\em local} Rossby number $Ro_{l}$ as a function of  radius for the hydrodynamical model H-Pr10-N5. $u$ and $l$ have been averaged over a few rotation periods.}
\label{fig:Rol}
\end{center}
\end{figure}

A more refined estimate for the rotational influence can be obtained via the {\em local} Rossby number~\citep{christensen2006}, defined here as function of radius by $Ro_{l}=\langle u \rangle_{\theta,\phi,t}/(\Omega l)$ with the longitudinally, latitudinally, and time averaged velocity $\langle u \rangle_{\theta,\phi,t}$. The length scale $l=\pi/\bar{\ell}_{u}$ is a characteristic length scale of the flow at radius $r$ defined using 
\begin{gather}
\bar{\ell}_{u} = \sum\limits_{\ell} \frac{\ell\langle\mathbf{u}_{\ell} \cdot \mathbf{u}_{\ell}\rangle_{\theta,\phi}}{\langle\mathbf{u} \cdot \mathbf{u}\rangle_{\theta,\phi}}
\end{gather}
where $\mathbf{u}_{\ell}$ is the flow component at a given radius for  spherical harmonic degree $\ell$. The radial variation of $Ro_{l}$ for model H-Pr10-N5 is shown in Fig.~\ref{fig:Rol}. Previous parameter studies have shown that as long as $Ro_{l}$ is smaller than a threshold value of $\approx 0.1$ Coriolis forces significantly affect the nature of convection~\citep{christensen2006, schrinner2012, gastine2013b, gastine2014}. Here $Ro_{l}<0.1$ in the entire shell. This explains why Fig.~\ref{fig:hydro_flow}({\bf a}) shows north-south aligned convection cells even near the outer boundary of the simulation.

\subsection{Self-consistent dynamos}

\subsubsection{Model NA-Pr10-N5}
\begin{figure*}
\begin{center}
\includegraphics[scale=0.44]{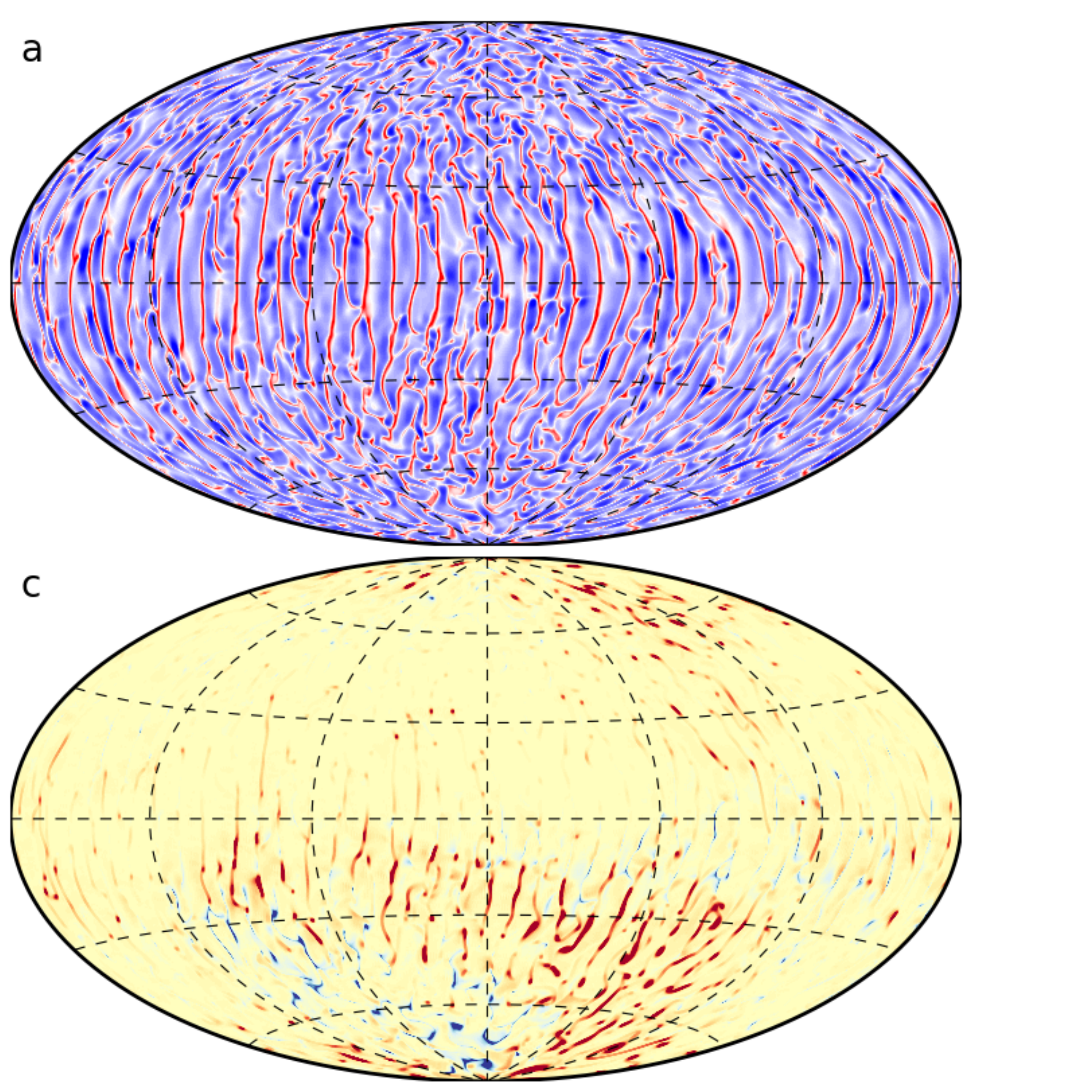}
\includegraphics[scale=0.44]{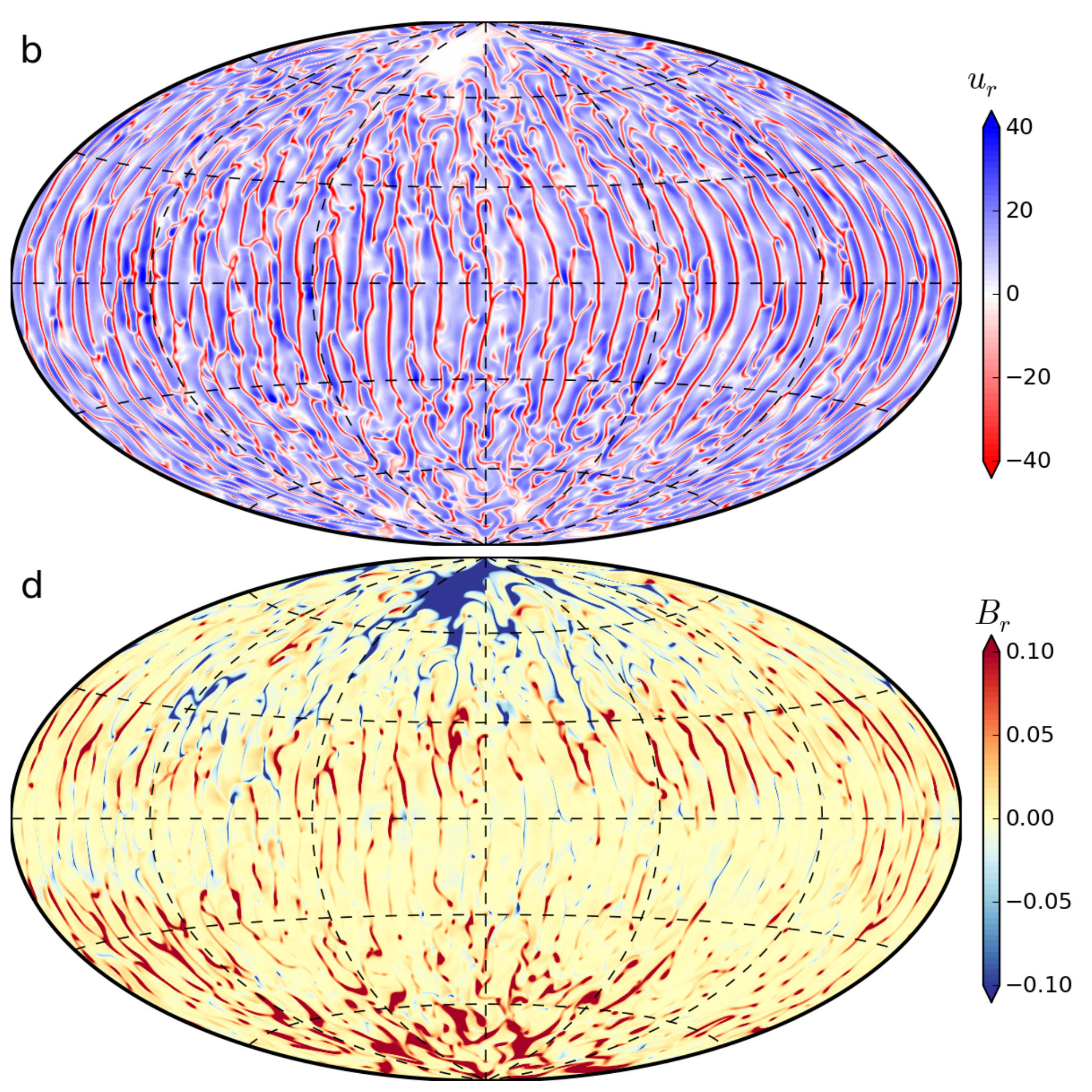}\\
\includegraphics[scale=0.44]{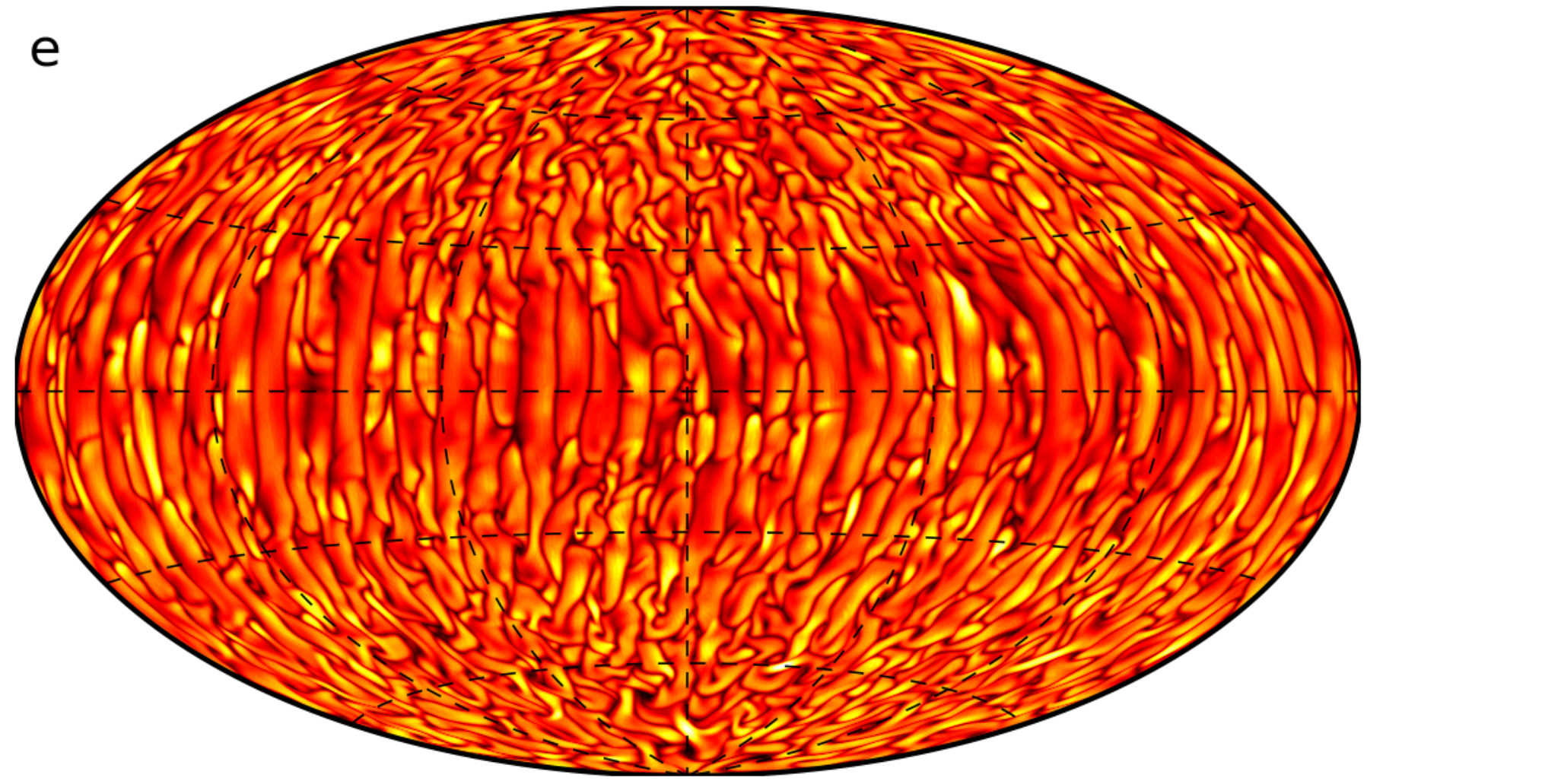}
\includegraphics[scale=0.44]{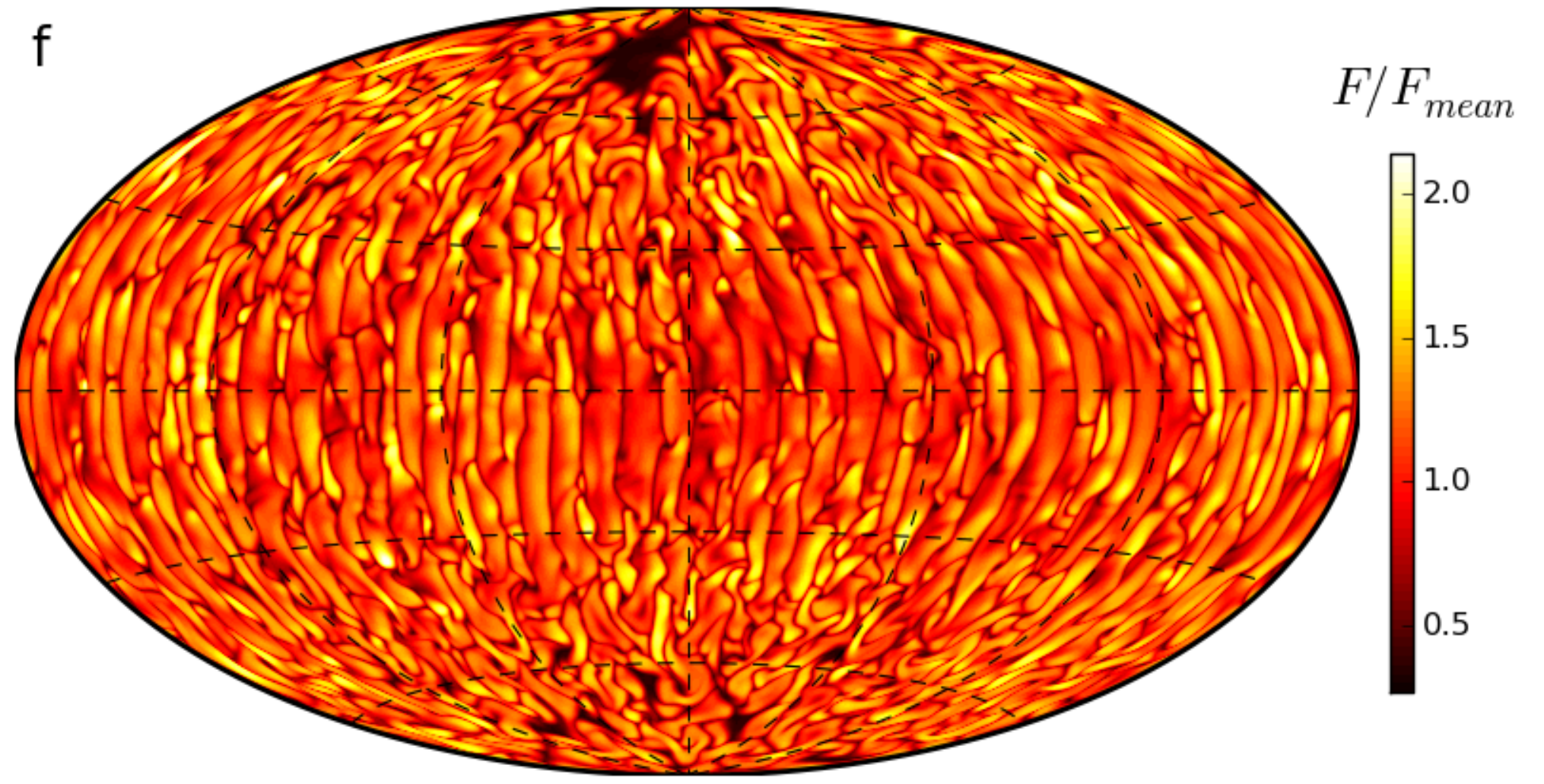}
\caption{Radial velocity $u_{r}$ at $r$=$0.99\,r_{o}$ in ({\bf a}), ({\bf b}), radial magnetic field $B_{r}$ at $r$=$0.99\,r_{o}$ in ({\bf c}), ({\bf d}), and total heat flux $F$ normalized by its surface mean $F_{mean}$ on outer surface in ({\bf e}), ({\bf f}) for model NA-Pr10-N5 and A-Pr10-N5 respectively.  $u_{r}$ is given in terms of the Reynolds number and the radial magnetic field is normalized by the equipartition field strength calculated using the time averaged total kinetic energy in the shell. The color variations for $u_r$ and $B_r$ are saturated at values lower than the extrema; the full variation range for $B_r$ is $\approx\pm0.3$ in ({\bf d}).}
\label{fig:Pr10_dyn_vr_Br}
\end{center}
\end{figure*}

We now turn to the dynamo action of the case discussed above. We use the hydrodynamic model H-Pr10-N5 as starting point and set $Pm$=10 which results in model NA-Pr10-N5. A very small seed magnetic field is exponentially enhanced by the helical convection and the system finally saturates to a statistically stationary state. A snapshot of the simulation in the saturated regime showing the radial velocity and the radial magnetic field near the outer boundary is shown in Fig.~\ref{fig:Pr10_dyn_vr_Br}({\bf a}) and ({\bf c}) respectively. The dynamo generated magnetic field is non-axisymmetric and the morphology is dominated by a spherical harmonic order $m$=1 component. The magnetic field at the instant shown in Fig.~\ref{fig:Pr10_dyn_vr_Br}({\bf c}) is also concentrated in the southern hemisphere. However, the hemisphere with dominant magnetic field can change with time~\citep{grote2000}. The field also propagates westwards in the rotating frame of reference of the shell. A "butterfly" diagram (not shown) of the azimuthally averaged radial field also shows poleward propagating features. Such travelling non-axisymmetric dynamo solutions have already been observed in dynamos with free-slip boundary conditions~\citep{schrinner2011, schrinner2012, kapyla2013} and can be described in terms of the classical Parker-waves~\citep{busse2006, goudard2008, schrinner2012}.

As shown in Fig.~\ref{fig:Pr10_dyn_vr_Br}({\bf c}) the divergent upwellings sweep the magnetic flux and concentrated it into the convergent downwellings. This sort of redistribution of magnetic flux is a generic trait in magnetic convection~\citep{proctor1982, vogler2005, stein2006, stein2012}. Furthermore, the expelled magnetic flux preferably accumulates into the nodes of the downwelling lanes of convection cells~\citep{stein2012}.

In Fig.~\ref{fig:Pr10_dyn_vr_Br}({\bf e}) we plot the total heat flux  normalized by its surface averaged value on the outer boundary of the simulation. The heat flux at the outer boundary is calculated as $F=-Pr^{-1}\tilde{\rho}\,\tilde{T}\,(ds/dr)$ where $\tilde{\rho}$ and $\tilde{T}$ are local (background) density and temperature, respectively, and $ds/dr$ is the local radial derivative of entropy. Comparing Fig.~\ref{fig:Pr10_dyn_vr_Br}({\bf a}) with Fig.~\ref{fig:hydro_flow}({\bf a}) shows that the radial flow is very similar in both cases. Nonetheless, a careful inspection of the nodes of the downwelling lanes does show a quenching of radial flow where the magnetic flux is concentrated. Correspondingly, the convective heat flux is also reduced in such regions (e.g. the tiny magnetic flux patch near the south pole in Fig.~\ref{fig:Pr10_dyn_vr_Br}({\bf c})).

\begin{figure}
\begin{center}
\includegraphics[scale=0.28]{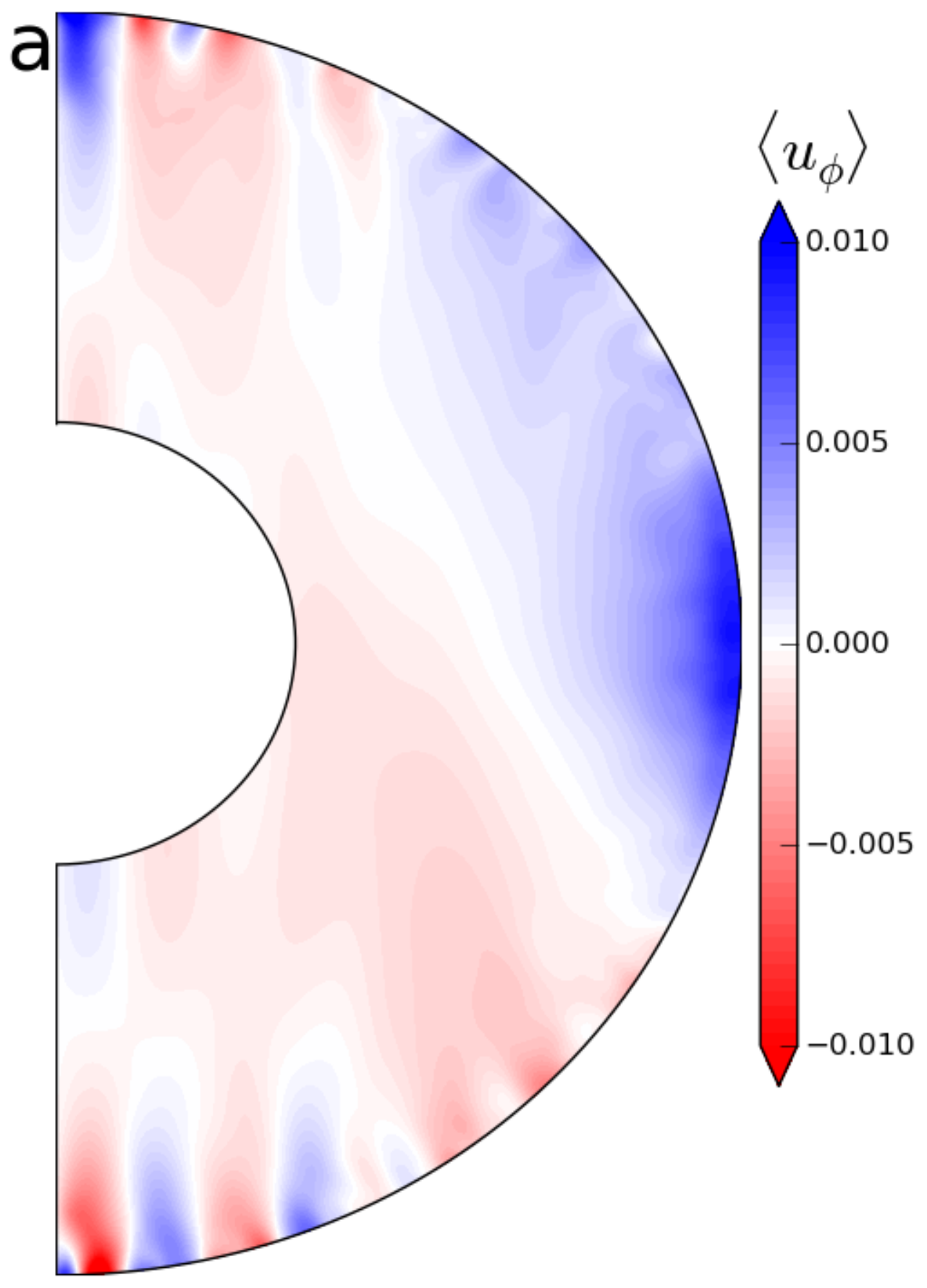}
\includegraphics[scale=0.36]{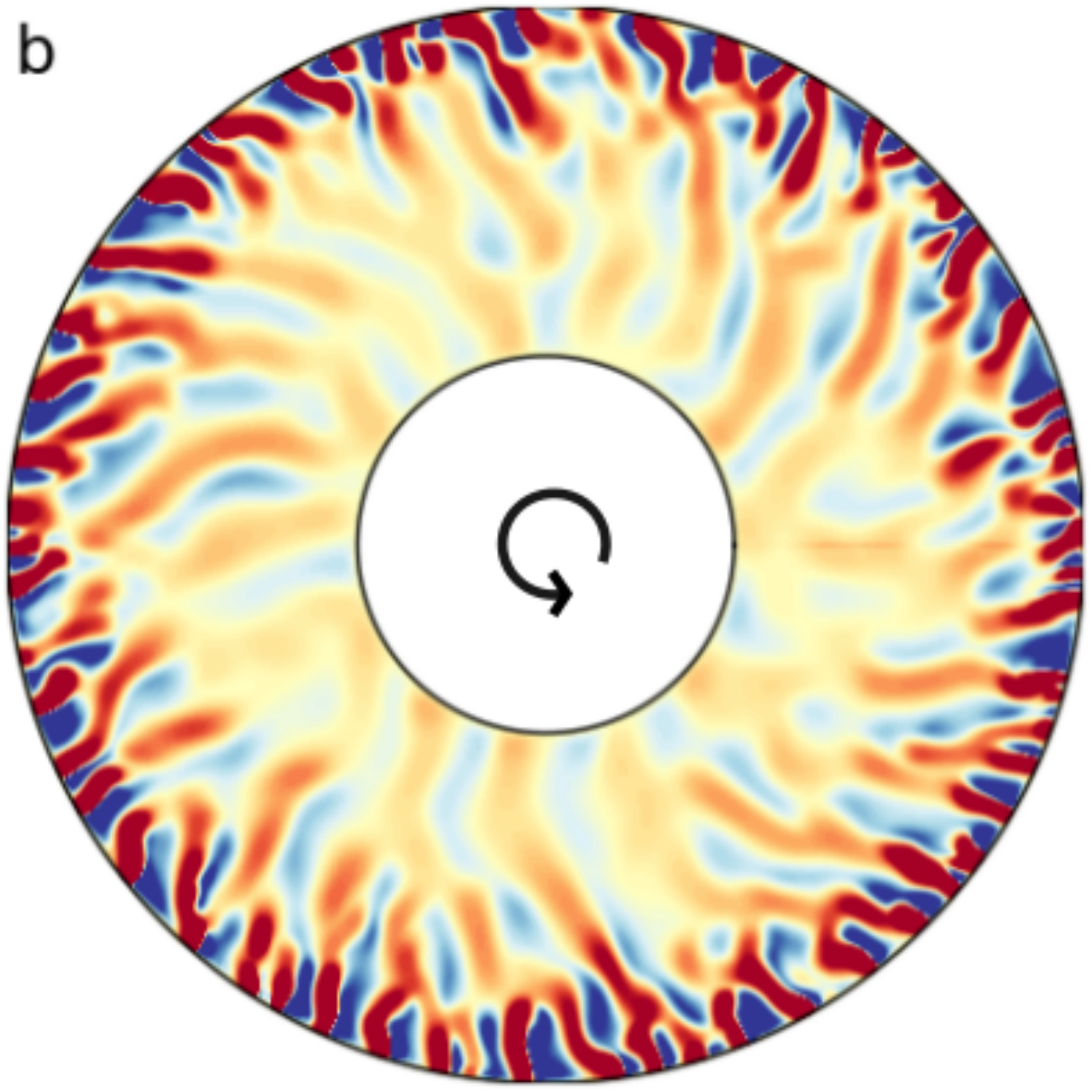}\\
\includegraphics[scale=0.28]{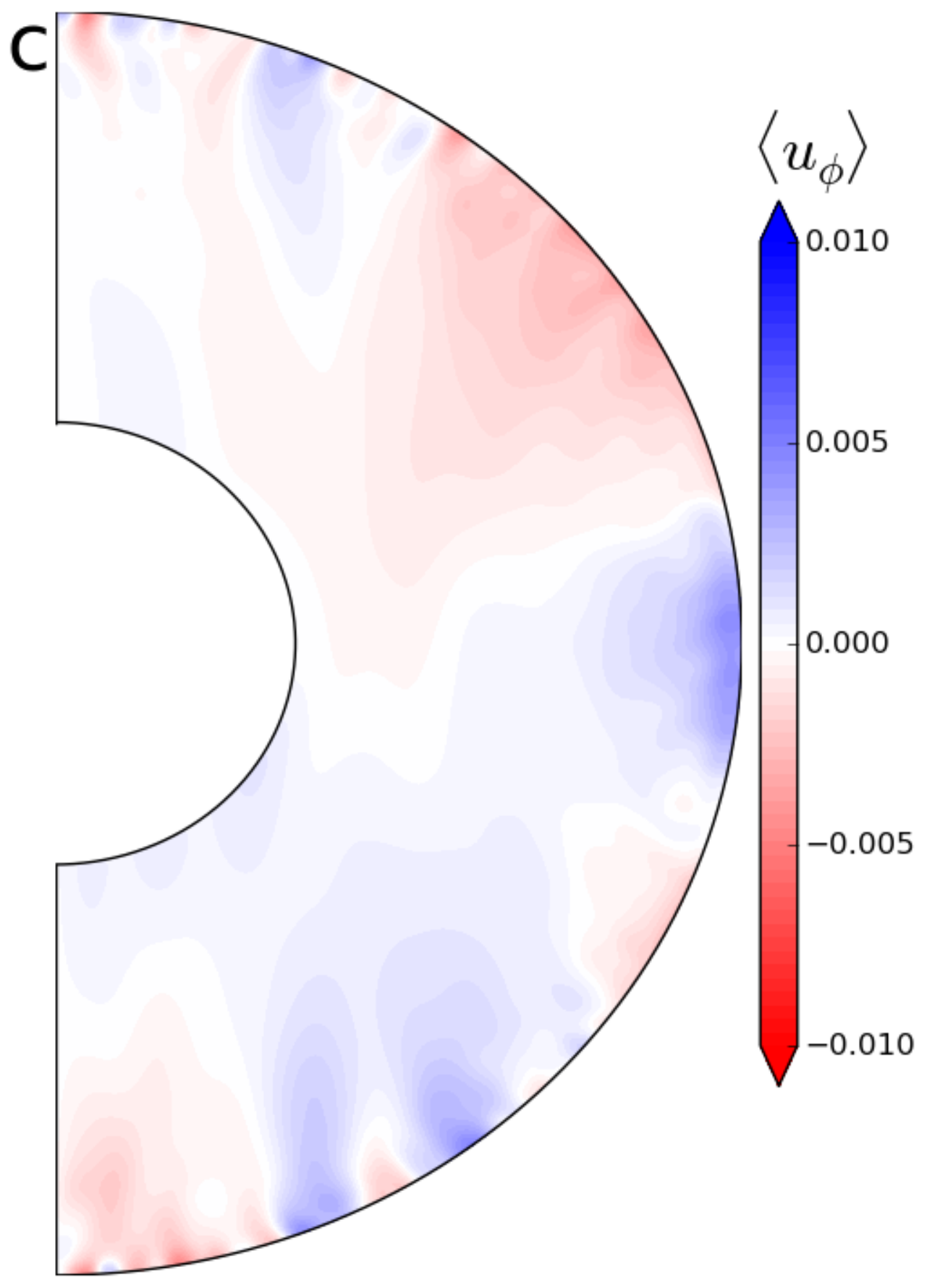}
\includegraphics[scale=0.17]{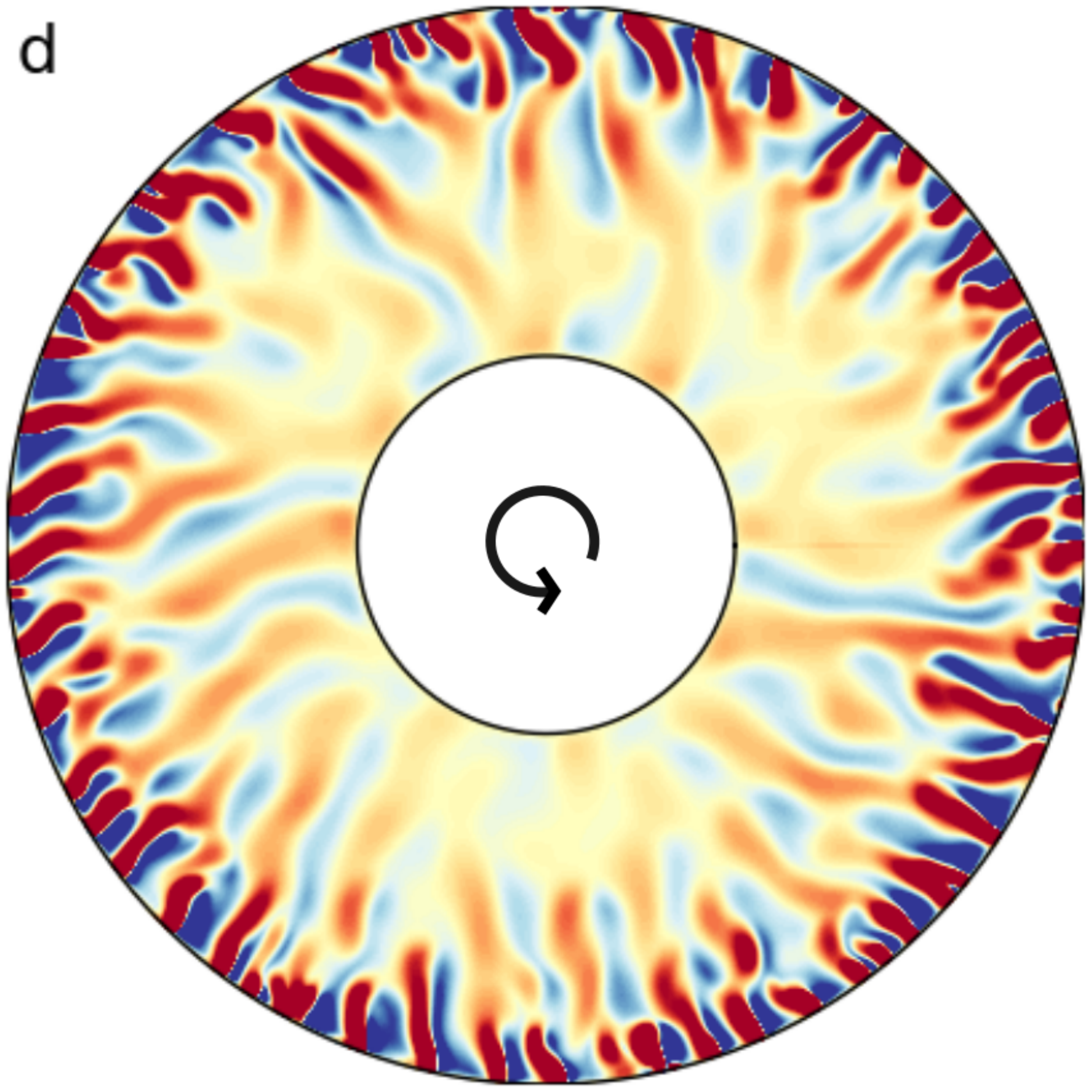}
\caption{Longitudinally averaged azimuthal flow in ({\bf a}) and ({\bf c}) and $\omega_z$ in ({\bf b}) and ({\bf d}) for model NA-Pr10-N5 and A-Pr10-N5 respectively.}
\label{fig:Pr10_dyn_zonal}
\end{center}
\end{figure}

\begin{figure}
\begin{center}
\includegraphics[scale=0.45]{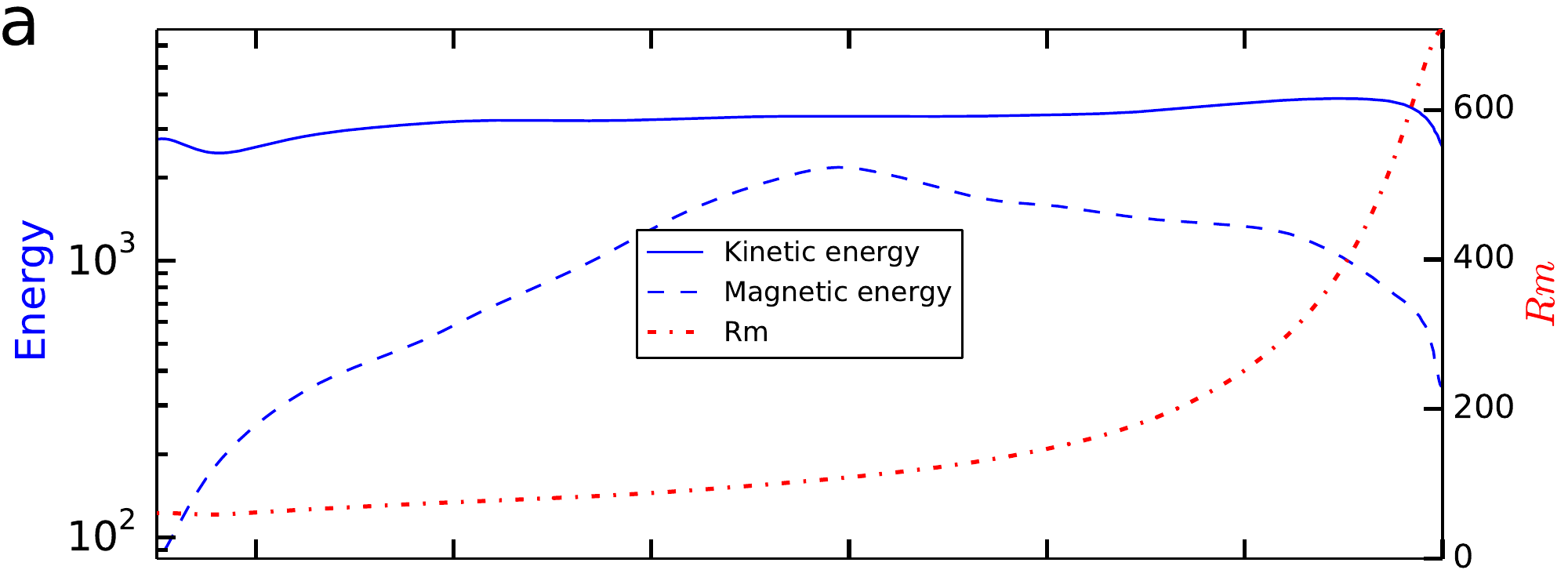}\\
\includegraphics[scale=0.45]{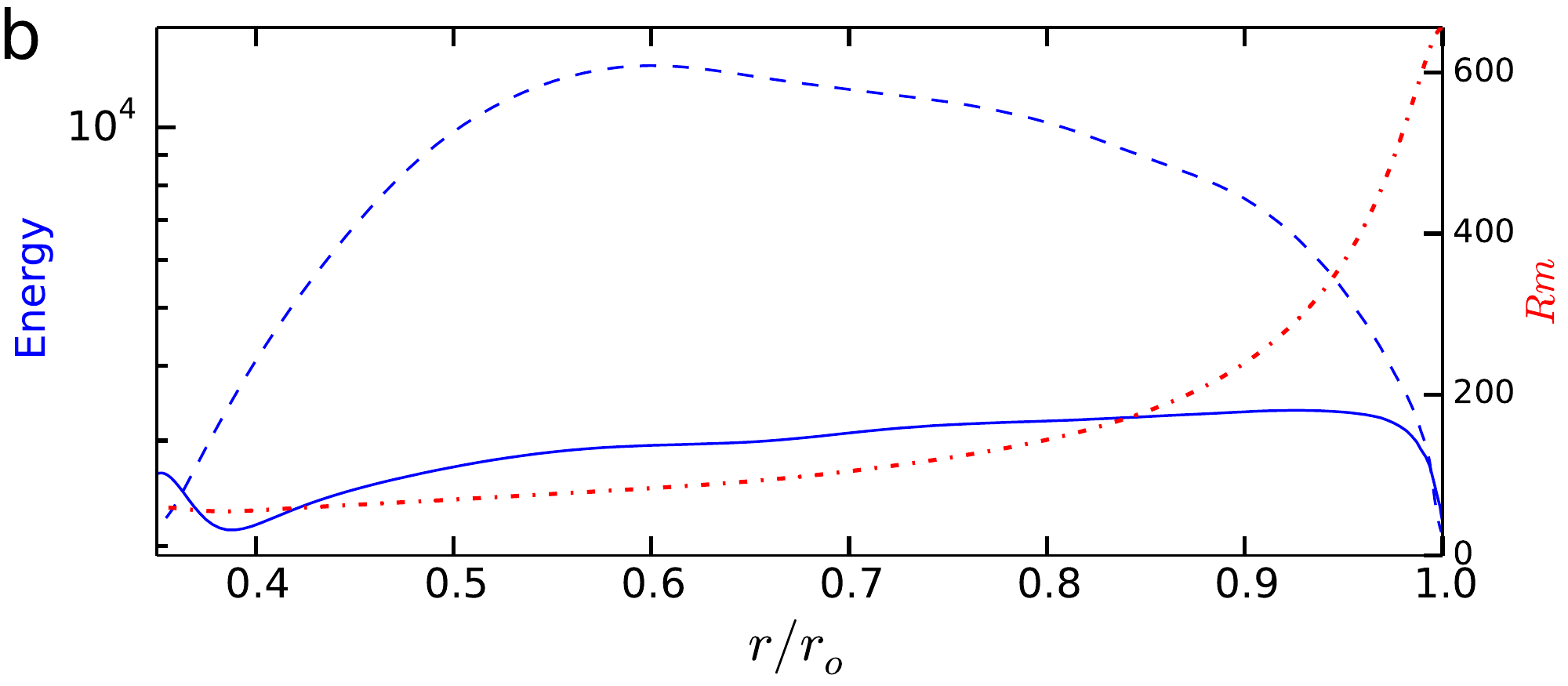}
\caption{Longitudinally and latitudinally averaged kinetic energy and magnetic energy density on the left axis, and magnetic Reynolds number $Rm=u\,d/\lambda$ on the right axis as a function of radius for model NA-Pr10-N5 (panel ({\bf a})) and A-Pr10-N5 (panel ({\bf b})). The quantities were averaged over few rotation periods.}
\label{fig:Pr10_dyn_energy}
\end{center}
\end{figure}

As compared to model H-Pr10-N5, the differential rotation (Fig.~\ref{fig:Pr10_dyn_zonal}({\bf a})) is quenched and the eastward tilt of helical convection columns (Fig.~\ref{fig:Pr10_dyn_zonal}({\bf b})) is reduced by the presence of the magnetic field in this case. The energy content of the axisymmetric differential rotation for H-Pr10-N5 and NA-Pr10-N5 is about 18\% and 8\%, respectively, of the total kinetic energy (Tab.~\ref{tab:tab1}). The qualitative structure of differential rotation, however, is similar in both cases (see Fig.~\ref{fig:hydro_flow}({\bf b}) and Fig.~\ref{fig:Pr10_dyn_zonal}({\bf a})). The total magnetic energy of model NA-Pr10-N5 is only about 30\% of the total kinetic energy (see Tab.~\ref{tab:tab1} and Fig.~\ref{fig:Pr10_dyn_energy}({\bf a})). In the mean-field formulation, such non-axisymmetric multipolar dynamos can be categorized as $\alpha\Omega$ or $\alpha^{2}\Omega$ type~\citep{schrinner2012, gastine2012b} where magnetic field co-exists with substantial differential rotation. Dynamo simulations resembling this case (i.e. cases with significant density stratification and a multipolar magnetic field) have been frequently reported~\citep{miesch2005, browning2008, ghizaru2010, kapyla2012, gastine2012b, schrinner2012, nelson2013, fan2014, cole2014} and do not exhibit any prominent low heat flux regions linked to strong magnetic fields that could be associated with starspots (see Fig.~\ref{fig:Pr10_dyn_vr_Br}({\bf e})).

\subsubsection{Model A-Pr10-N5: Polar starspots}
Instead of initiating the dynamo simulation with a tiny seed magnetic field we now take the flow input from model H-Pr10-N5 and initiate the dynamo simulation with strong dipolar magnetic field aligned with the rotation axis. The resulting dynamo model is A-Pr10-N5 (Tab.~\ref{tab:tab1}). The motivation behind starting with a strong dipolar field is to have strong Lorentz forces that can quench the shearing differential-rotation via Maxwell stresses~\citep{ferraro1937, aubert2005}. This allows an axial-dipole dominant (ADD) solution to develop and stabilize. As shown in Fig.~\ref{fig:Pr10_dyn_zonal}({\bf c}) the axisymmetric differential rotation becomes even more strongly quenched (energy content about 3\% of total kinetic energy) as compared to the non-magnetic model than in case NA-Pr10-N5 and and has lost entirely any semblance to geostrophy. Reynolds stresses are not effective any more as the spiralling of convection columns nearly vanishes (Fig.~\ref{fig:Pr10_dyn_zonal}({\bf d})). Such differential rotation profile is typically associated with a thermal wind balance~\citep{aubert2005}, i.e. a balance of buoyancy and Coriolis forces. This ADD solution is stable as we have run this simulation for about 2 magnetic diffusion time ($d^2/\lambda$). This ADD configuration would have decayed after a magnetic diffusion time if it was not self-consistently sustained by the convection. Note that one magnetic diffusion time is equal to one thermal diffusion time ($d^2/\kappa$) since both $Pm$ and $Pr$ are equal for this case. Therefore, the simulation is also thermally relaxed.

Dynamos with ADD magnetic field that co-exist with highly quenched  differential rotation are classified as $\alpha^{2}$-dynamos~\citep{olson1999, chabrier2006, schrinner2012}. Dynamos in this state are said to be in a "magnetostrophic" state where Lorentz forces are rather strong and enter in the first order force balance \citep[e.g.][]{sreenivasan2006}. Generally, parameter studies~\citep{browning2008, gastine2012b, yadav2013a} show that large scale magnetic fields generated by a distributed dynamo quench the differential rotation to low values (much lower than the solar case). Recently, this could be empirically verified for a large sample of cool stars which show a rough inverse correlation between rotation rate and differential rotation~\citep{reinhold2013}. Note that higher rotation rates generally imply stronger magnetic fields in cool stars~\citep{pizzolato2003}.

\begin{figure}
\begin{center}
\includegraphics[scale=0.48]{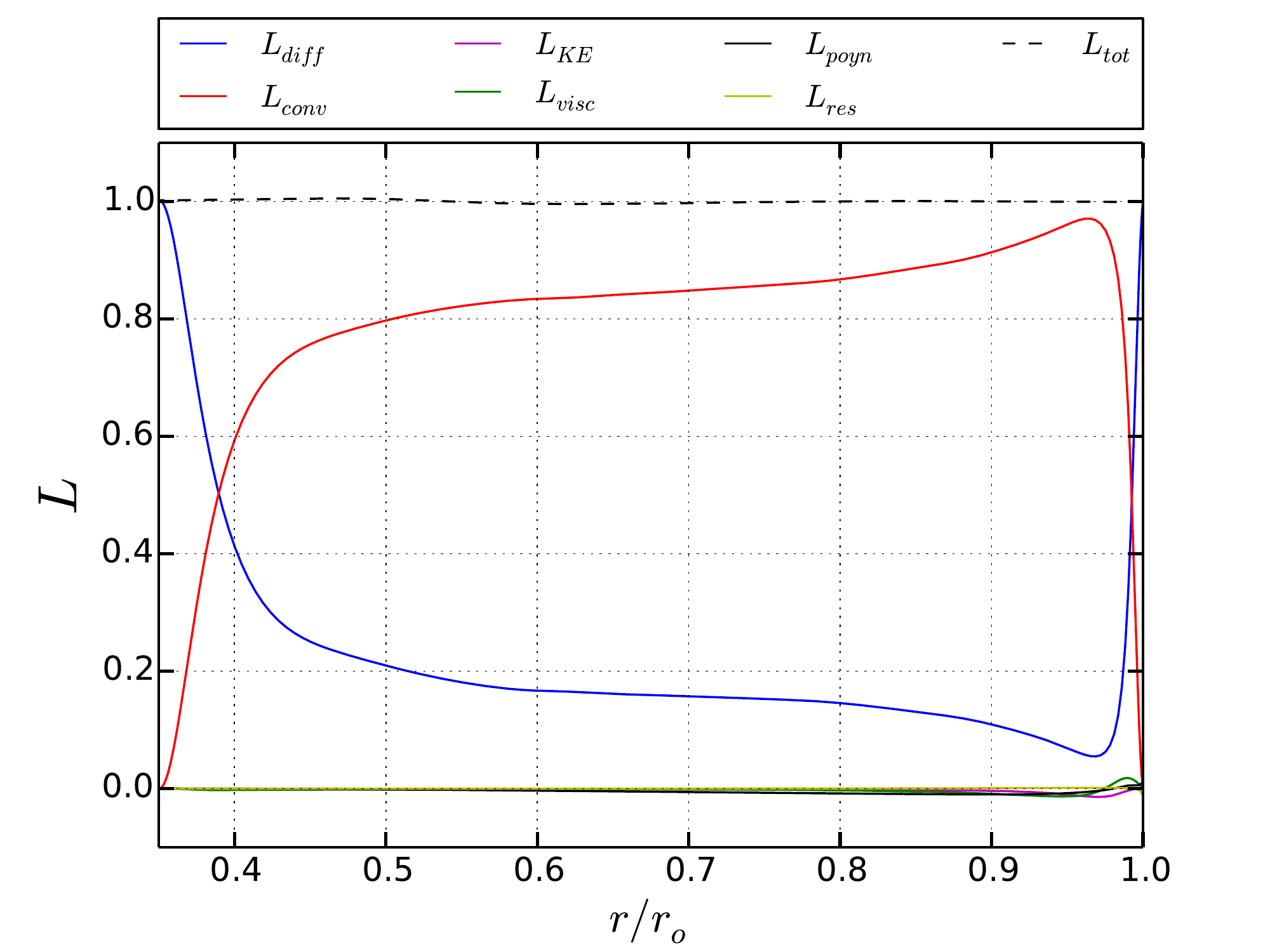}
\caption{Radial profiles of non-dimensional luminosities, i.e. $r^2\int{\langle F\rangle_t sin(\theta)\,d\theta\, d\phi}$, associated with the different fluxes for model A-Pr10-N5, with $L_{tot}$ being the sum of all contributions. The different contributions are normalized by $L_{tot}$ at $r=r_o$. The profiles were averaged for about 200 rotations.}
\label{fig:Pr10_dip_rad_flux}
\end{center}
\end{figure}

Figure~\ref{fig:Pr10_dip_rad_flux} displays the radial profile of the non-dimensional luminosity $L$ calculated using the various energy fluxes defined by Eqs.~(\ref{eq:flux1}-\ref{eq:flux6}). The total luminosity $L_{tot}$ is nearly constant throughout the convection zone, signifying the statistically stationary nature of the solution. Conductive and convective contributions dominate the energy transport, with the former dominating near the boundaries while the latter in the bulk. Assuming that the thermal boundary layers (or rather `entropy' boundary layers since entropy diffusion is assumed in our formulation) extend up to the point where diffusive and convective flux contributions are equal \citep[see e.g.][]{julien2012}, the bottom and top boundary layers span about $0.04\,r_o$ and $0.01\,r_o$, respectively. The thickness difference between these two boundary layers can be attributed to the large density contrast in the simulated convection zone. Viscous, Poynting, and resistive fluxes are only secondary contributions, similar to what is usually observed in such global convection simulations. Owing to the relatively small role of inertia in this dynamo model (since $Pr$=10) the kinetic energy flux also makes a small contribution as compared to earlier studies which frequently employed $Pr\le\,1$ \cite[e.g.][]{miesch2005, nelson2013, fan2014}.

A snapshot of radial velocity and radial magnetic field near the outer boundary is given in Fig.~\ref{fig:Pr10_dyn_vr_Br}({\bf b}) and ({\bf d}) respectively. Unlike the multipolar dynamo model NA-Pr10-N5, the quenching of near-surface flows in regions of strong magnetic field is rather prominent in this case and is clearly visible in Fig.~\ref{fig:Pr10_dyn_vr_Br}({\bf b}), especially at high latitudes. The size of the big patch near the north pole with very weak radial flow is larger than the general length scale of the convection cells. At low latitudes convection forms irregular north-south aligned lanes which are associated with narrow elongated flux concentrations that are relatively short-lived. Here the field concentration is usually not strong enough to seriously impede radial flow. We use the term `dark spot' for a region of substantial size (similar to or larger than the local convection cells) on the surface of the model where the heat flux has been suppressed by  at least $\approx$50\% below the average surface heat flux. As locally very strong magnetic fields severely quench the flow, the convective heat transport is reduced which leads to the formation of dark spots that can be associated with cool starspots.

A comparison of the radial variation of kinetic and magnetic energy of model A-Pr10-N5 (Fig.~\ref{fig:Pr10_dyn_energy}({\bf b})) with that of NA-Pr10-N5  (Fig.~\ref{fig:Pr10_dyn_energy}({\bf a})) reveals that the magnetic energy dominates in the entire convection zone (on average) in the case with dipolar magnetic field, i.e. this case generates super-equipartition magnetic field. Note that the production of super-equipartition fields is not novel and geodynamo simulations frequently produce such strong magnetic fields, mimicking the scenario in Earth where magnetic energy is thought to be much larger than the kinetic energy (by a factor of about 7000)~\citep{roberts2000}. In the stellar context \citet{featherstone2009} have reported a spherical shell dynamo with super-equipartition field strengths.  Systematic numerical studies have shown that  dipolar dynamos in general produce higher field strengths than the multipolar ones~\citep{christensen2010, schrinner2012, gastine2012b, yadav2013a, yadav2013b}.

\begin{figure}
\begin{center}
\includegraphics[scale=0.34]{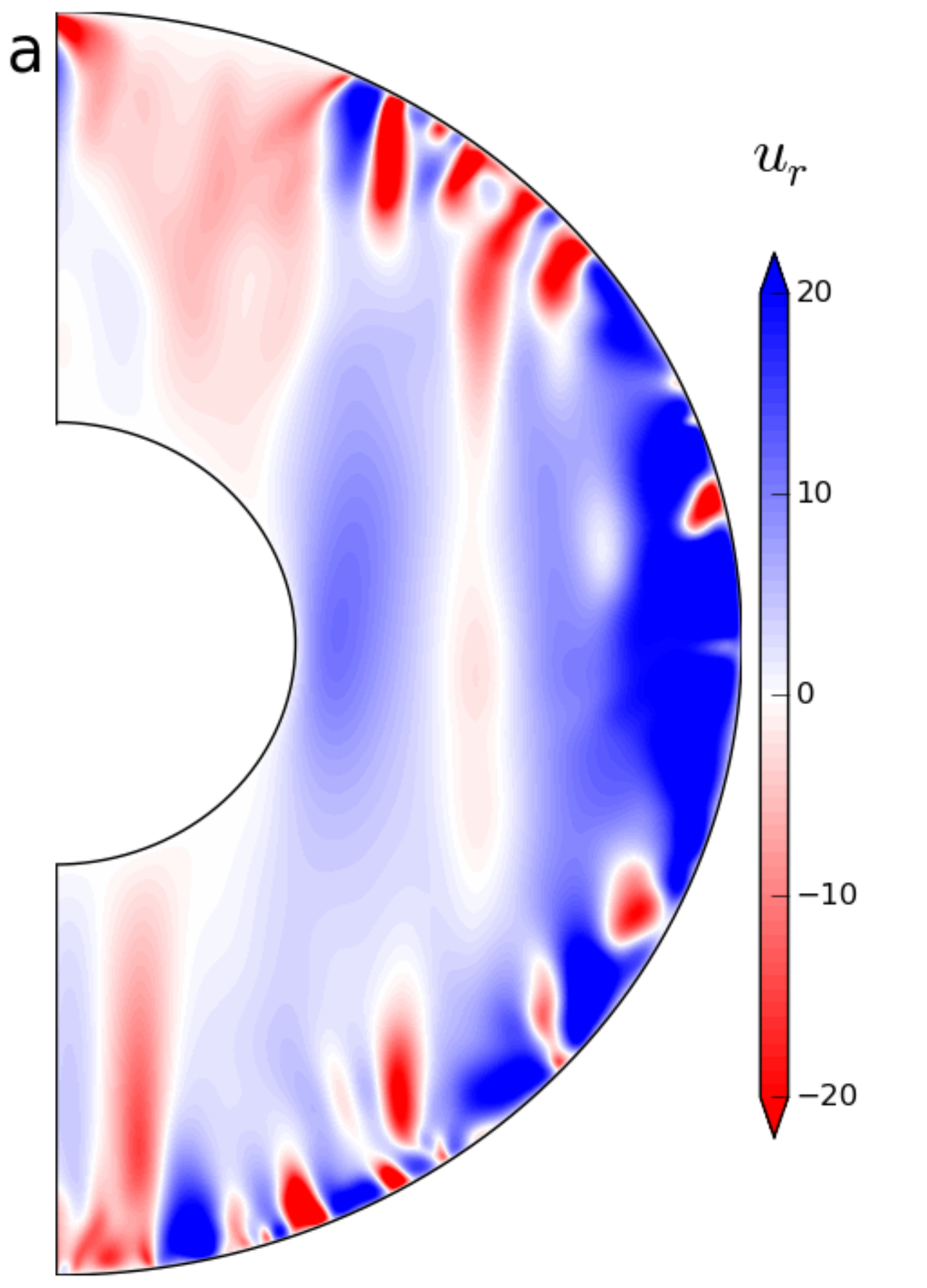}
\includegraphics[scale=0.34]{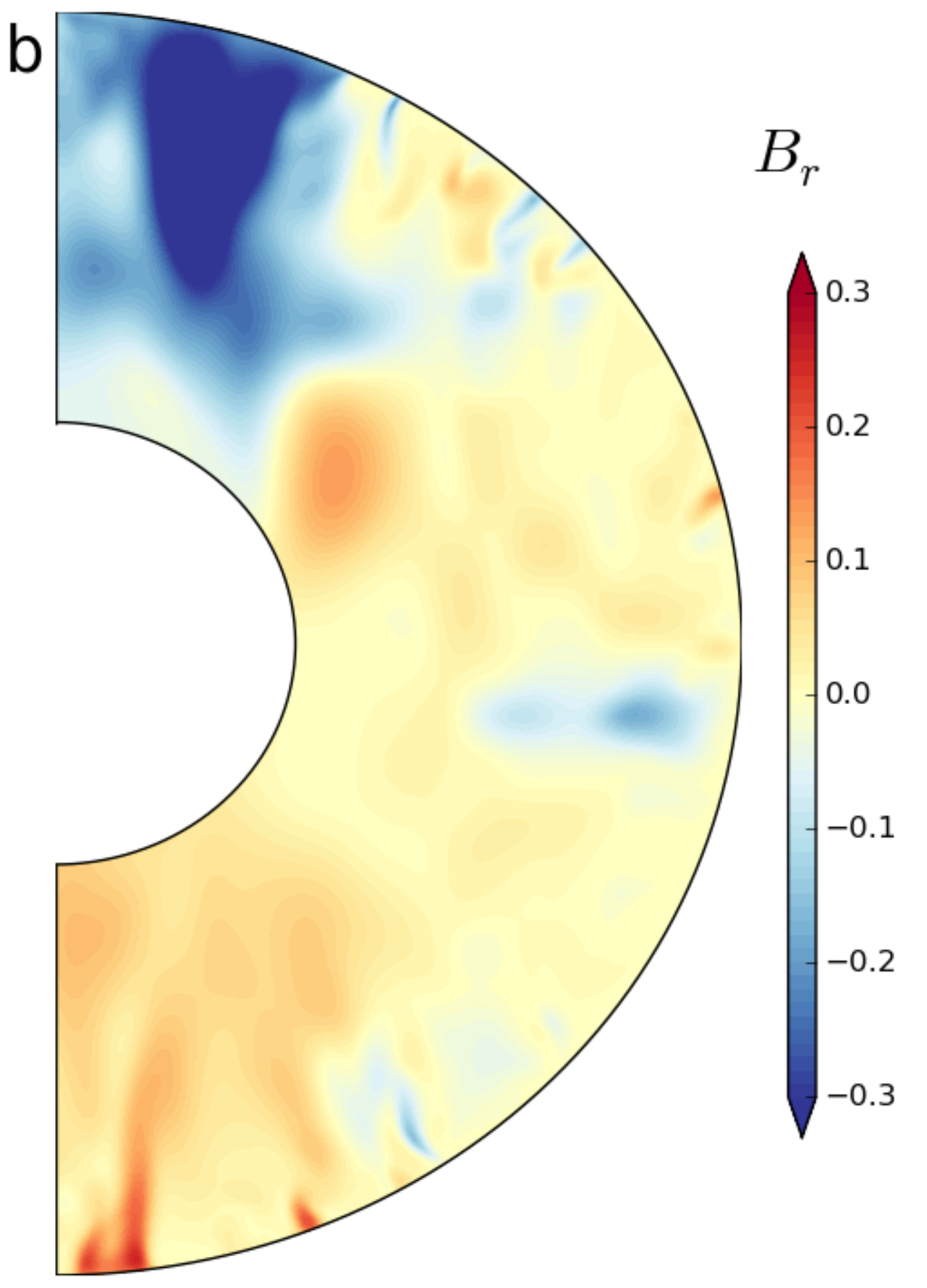}
\caption{Radial velocity in ({\bf a}) and radial magnetic field in ({\bf b}) on a constant longitude passing through the big polar spot for model A-Pr10-N5. The color scales saturate at values lower than the extrema.}
\label{fig:Pr10_dip_slice}
\end{center}
\end{figure}

The quenching of convective flow that leads to dark spot formation in this simulation is a relatively shallow phenomenon. For instance, the flow suppression at the instant shown in Fig.~\ref{fig:Pr10_dyn_vr_Br} is noticeable down to a depth of about $0.95\,r_o$. However, although  shallow, the quenching of convection extends well beyond the outer thermal boundary layer which reaches down to about $0.99\,r_o$. Figure \ref{fig:Pr10_dip_slice} shows the radial velocity and the radial magnetic field on a cut along the rotation axis at a longitude passing through the big northern spot in Fig.~\ref{fig:Pr10_dyn_vr_Br}({\bf d}). As is typical in ADD dynamos, a high concentration of magnetic flux exists at high latitudes. The magnetic field associated with this spot is deep seated. The integrity of this prominent flux-concentration seems to be maintained by rapid convective downwellings which surround it at its lateral margins (Fig.~\ref{fig:Pr10_dip_slice}({\bf a})).

\begin{figure*}
\begin{center}
\includegraphics[scale=0.255]{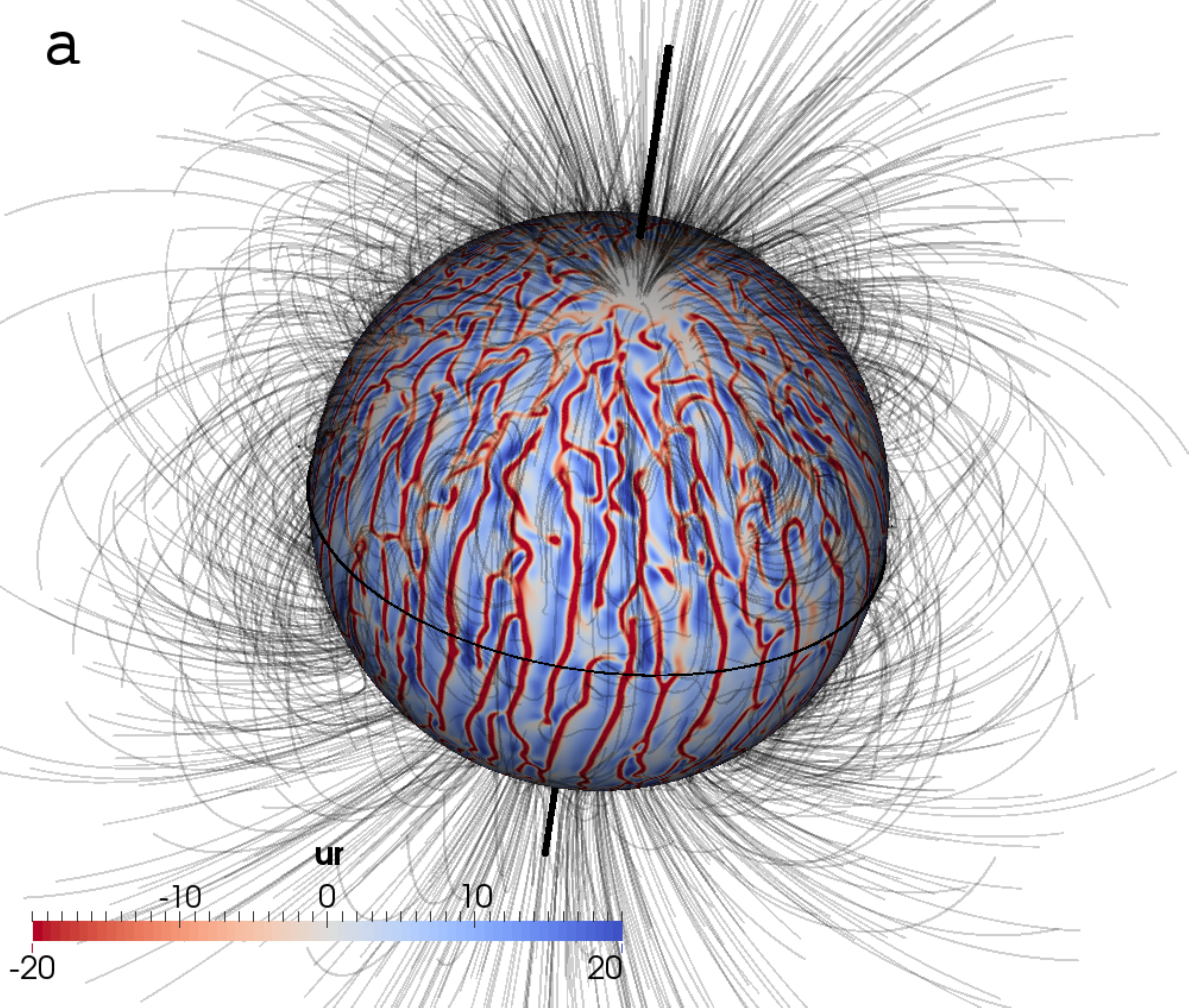} \includegraphics[scale=0.255]{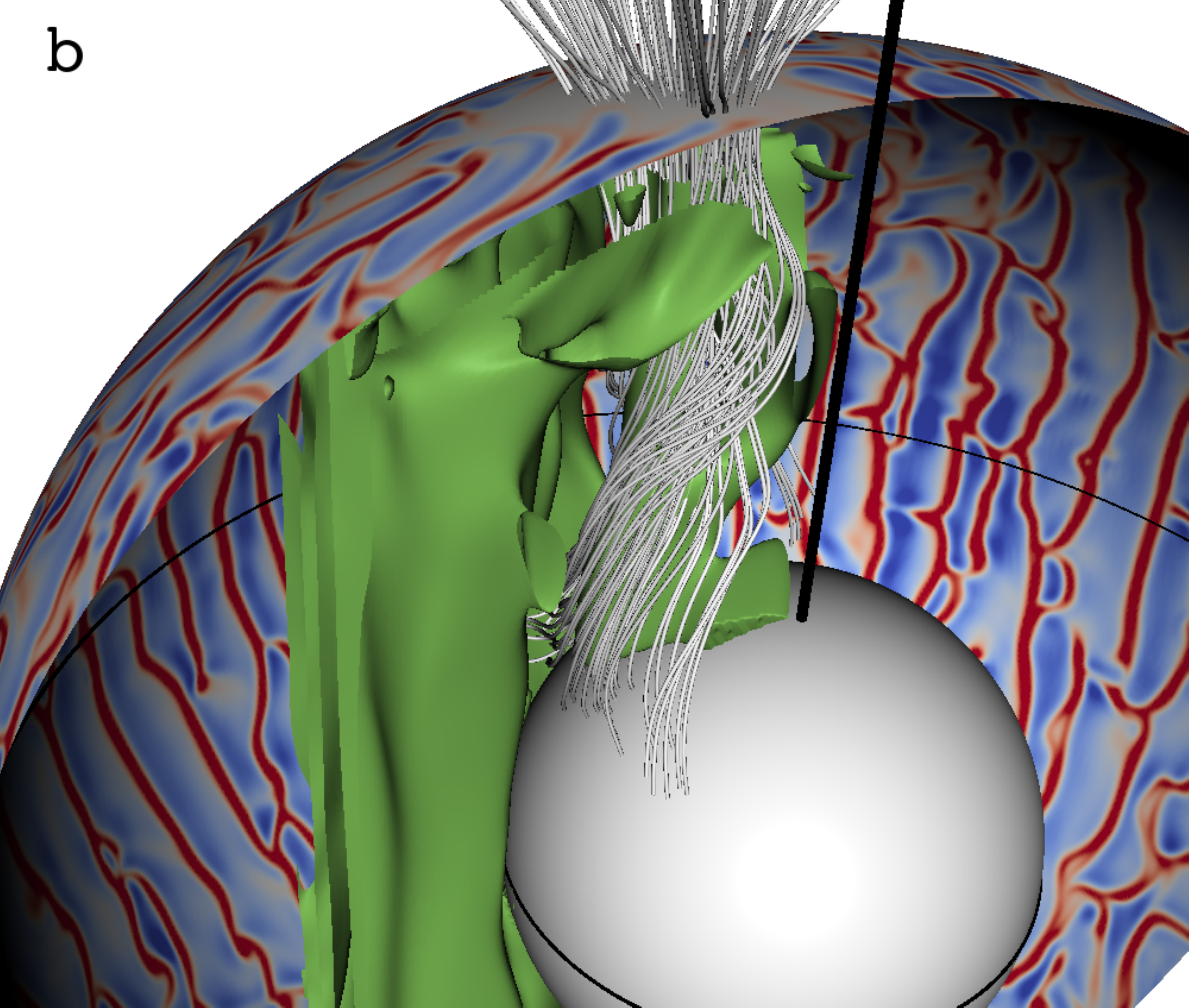} \\
\caption{Perspective view of model A-Pr10-N5 (polar spot case) from a northern latitude. ({\bf a}) radial velocity near the outer boundary and the magnetic field lines above the shell, upward continued by assuming a potential field. In ({\bf b}) the surface showing the radial flow in ({\bf a}) is cut and only the eastern hemisphere in shown in order to visualize the interior magnetic field lines associated with the northern big dark spot (visible as white patch in ({\bf a})). Isosurface of axial vorticity $\omega_z$ (in green, shown in a limited domain for clarity) are also illustrated in ({\bf b}).}
\label{fig:3D_render}
\end{center}
\end{figure*}

\subsubsection{Time evolution of darkspots}
The attached animation, "$\text{Br.mp4}$", shows the rich dynamics of the radial magnetic field on the {\em outer} boundary of the model A-Pr10-N5. The units in the animation are similar to Fig.~\ref{fig:Pr10_dyn_vr_Br}({\bf f}) and it spans about 75 rotations. It shows a prominent high-latitude flux-concentration which evolve on a much longer time scale than the local convection. Big flux-patches stay for many rotations and form when two or more sizeable patches merge. The elongated north-south aligned convection cells near the equator act as clear pathways for tiny flux patches to migrate latitudinally.

\begin{figure}
\begin{center}
\includegraphics[scale=0.45]{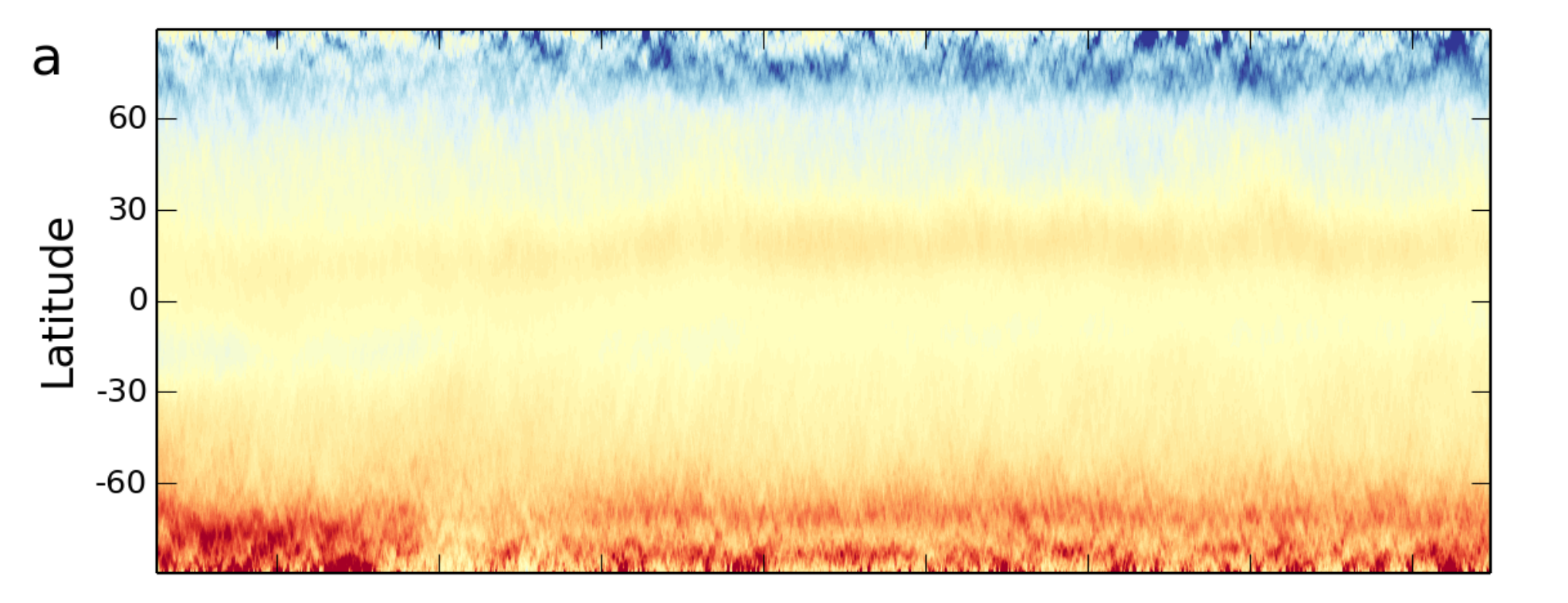}
\includegraphics[scale=0.45]{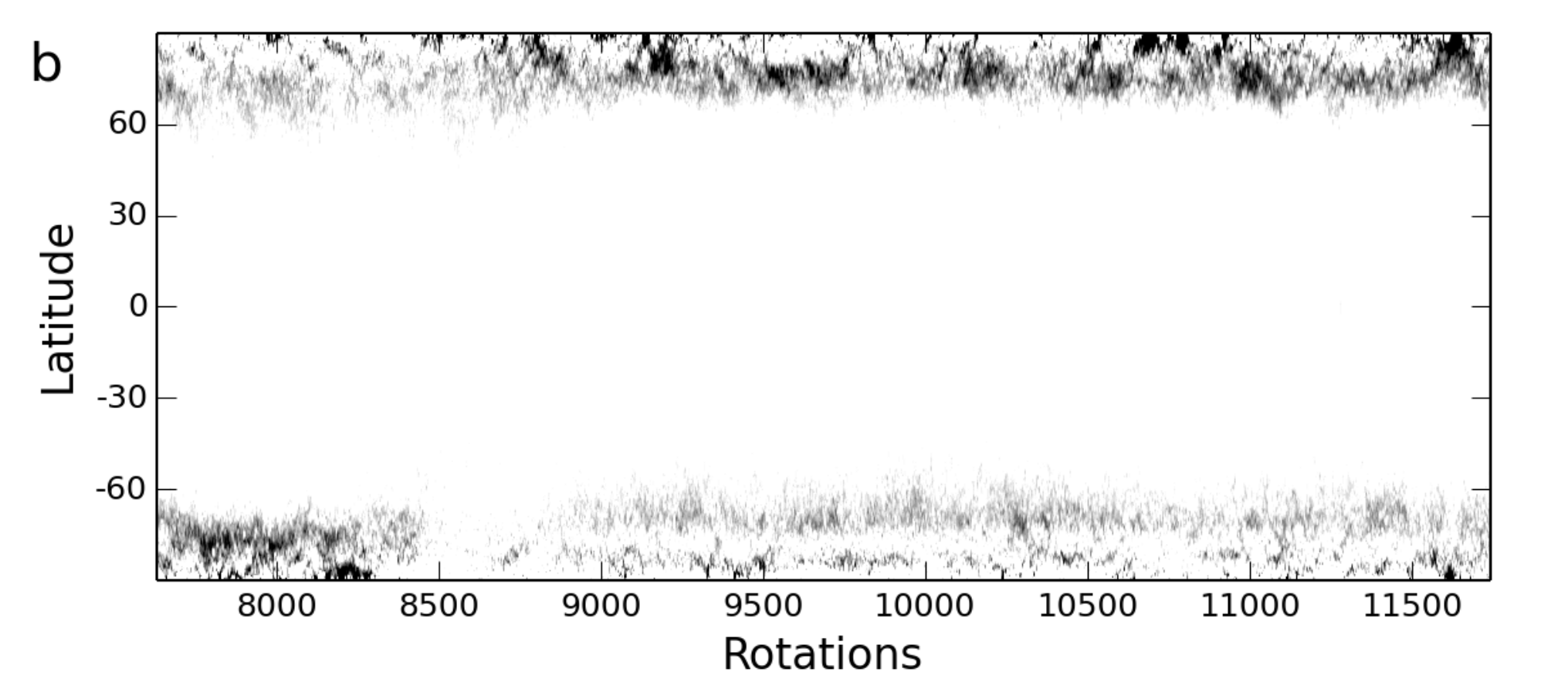}
\caption{Panel ({\bf a}) shows the azimuthally averaged radial magnetic field on the outer boundary of model A-Pr10-N5 (color map similar to Fig.~\ref{fig:Pr10_dyn_vr_Br}({\bf d})). Panel ({\bf b}) highlights magnetic field induced dark regions on the outer boundary. The latter was constructed by calculating the azimuthally averaged relative heat flux (sampled in Fig.~\ref{fig:Pr10_dyn_vr_Br}({\bf f})) for the simulation and plotting data which was less than unity.}
\label{fig:butterfly}
\end{center}
\end{figure}

Figure~\ref{fig:butterfly} provides a long term perspective on the simulation by displaying azimuthally-averaged radial magnetic field and relative heat flux as a function of time. Generally, the magnetic field shown in ({\bf a}) is dominated by an ADD configuration. However, the dynamo solution also portrays long term dynamics where the hemisphere with larger magnetic flux switches from one to the other as the simulations progresses (for instance, at around 9000 rotations). The change in the magnetic flux of a hemisphere is clearly visible in ({\bf b}) which highlights the evolution of the darkest regions on the outer boundary of the simulation. Here too, the hemisphere with darker spots switches from southern to northern hemisphere. Although low latitude regions do have small magnetic field induced spots, the azimuthal averaging filters them out, explaining the absence of any feature at low latitudes.

In Fig.~\ref{fig:3D_render}, we show a 3-D rendering of model A-Pr10-N5. Figure~\ref{fig:3D_render}({\bf a}) clearly shows the large scale component of the magnetic fields which is dominated by an axial dipole. Figure \ref{fig:3D_render}({\bf b}) shows that the field lines in the northern dark spot are rooted in deeper convection columns. Based on Fig.~\ref{fig:3D_render}({\bf b}) we can conjecture the following  formation mechanism for {\em sizeable} dark spots: first, helical columnar convection in the interior generates a collection of twisted and mainly radial magnetic field lines, and, second, the integrity of this stable magnetic structure is maintained by the downwellings of the convection in outer layers at its edges (see Fig.~\ref{fig:Pr10_dip_slice} as well). Since deep-rooted columns are wider and have a larger evolution time scale (sluggish velocity due to higher density)  the flux bundles associated with them would appear as a rigid structure to near surface convection. The dark spots formed due to these "anchored" flux bundles evolve on a longer time scale as compared to local convection.

\begin{figure}
\begin{center}
\includegraphics[scale=0.44]{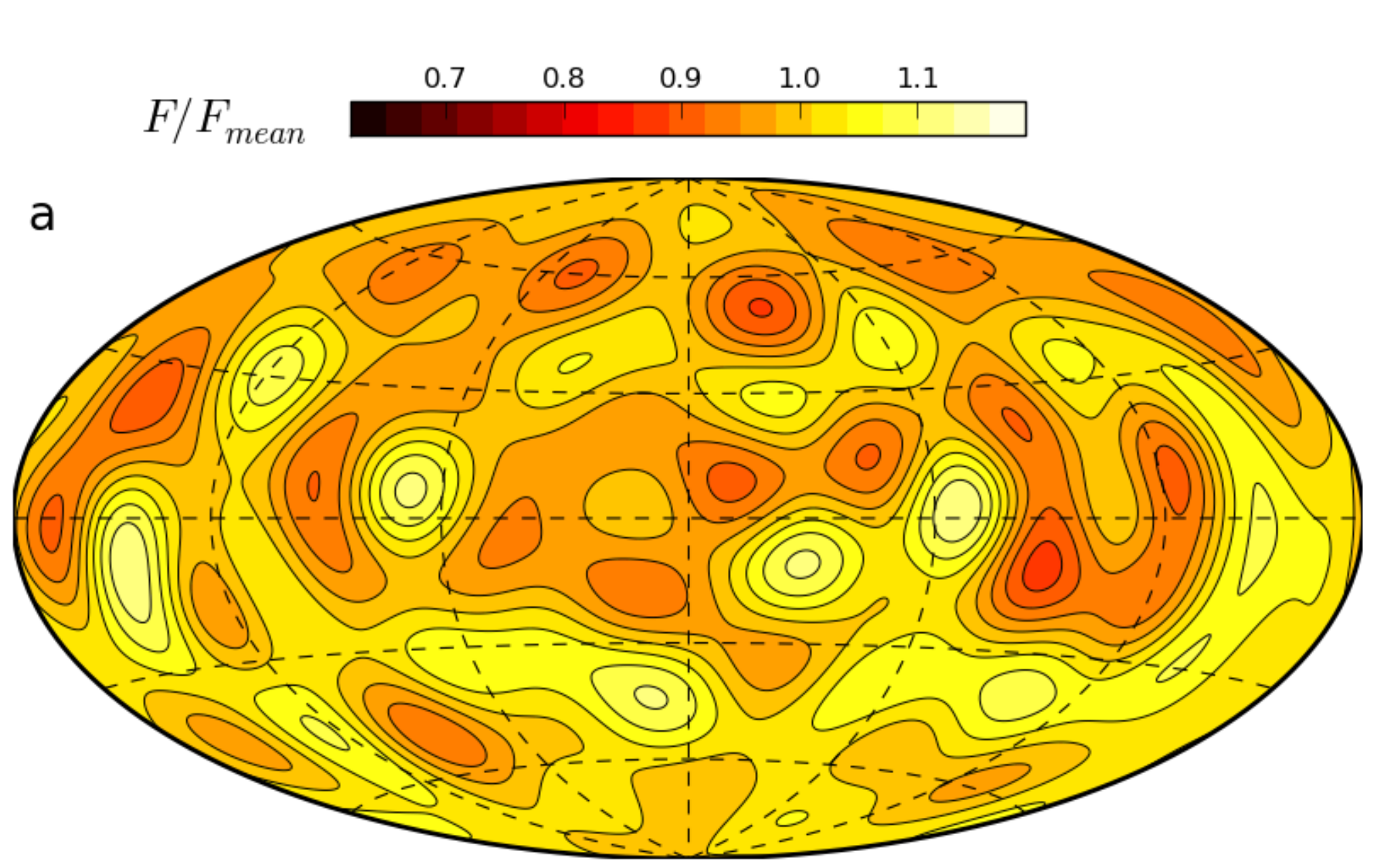}\\
\includegraphics[scale=0.44]{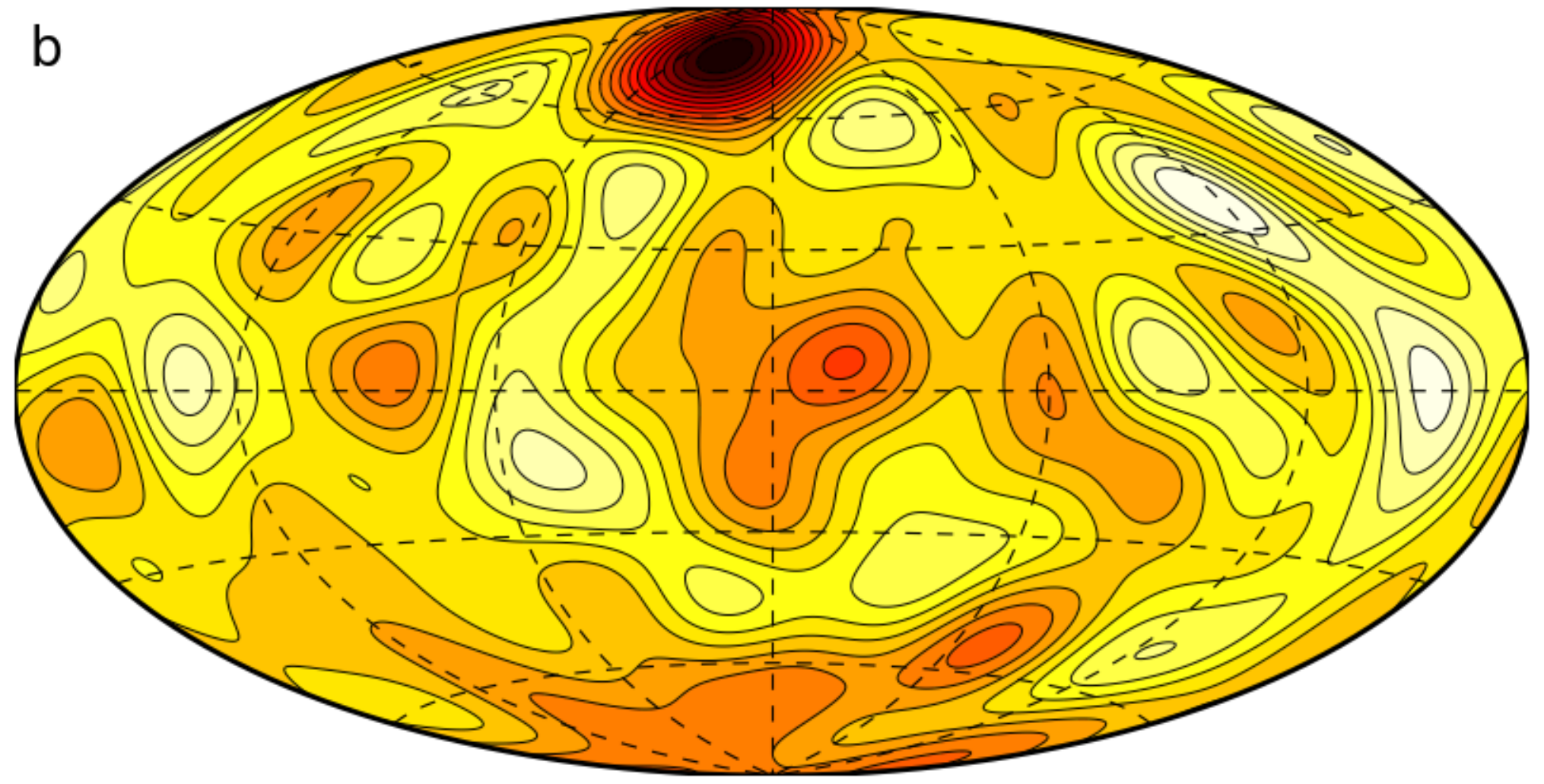}\\
\caption{Low-order representation of Fig.~\ref{fig:Pr10_dyn_vr_Br}({\bf e}) and Fig.~\ref{fig:Pr10_dyn_vr_Br}({\bf f}) in ({\bf a}) and ({\bf b}), respectively.}
\label{fig:Pr10_dip_flux_trunc}
\end{center}
\end{figure}

Observational techniques for starspot detection, such as Doppler imaging, provide only a smeared-out picture of the stellar surface. Therefore, the information about small scale features is washed out. The high latitude dark spots are usually assumed to be a collection of smaller dark spots~\citep{berdyugina2005, strassmeier2009}. To better compare our simulation results with the observational Doppler images we smear out the details in the simulations by filtering out all spherical harmonic degrees beyond 10 from the original simulation data. The resulting heat-flux fluctuation maps are shown in Fig.~\ref{fig:Pr10_dip_flux_trunc}. In the non-dipolar case NA-Pr10-N5 shown in ({\bf a}) the fluctuations are moderate and no consistent pattern exists. In the ADD case A-Pr10-N5 shown in ({\bf b}) the dark spot in the polar region has become even more prominent. Figure \ref{fig:Pr10_dip_flux_trunc} also shows other low-contrast features on the surface which do not represent magnetic field driven dark spots. The attached animation "$\text{LowOrder-Flux.mp4}$" shows the time evolution of low-order heat-flux fluctuations corresponding to animation "$\text{Br.mp4}$". The polar spots in the north pole region maintain a broad dark feature which persists throughout the animation. Other low-contrast features in the animation near the equator and near the south pole are more dynamic.

\subsubsection{Synthetic light curves}

\begin{figure}
\begin{center}
\includegraphics[scale=0.46]{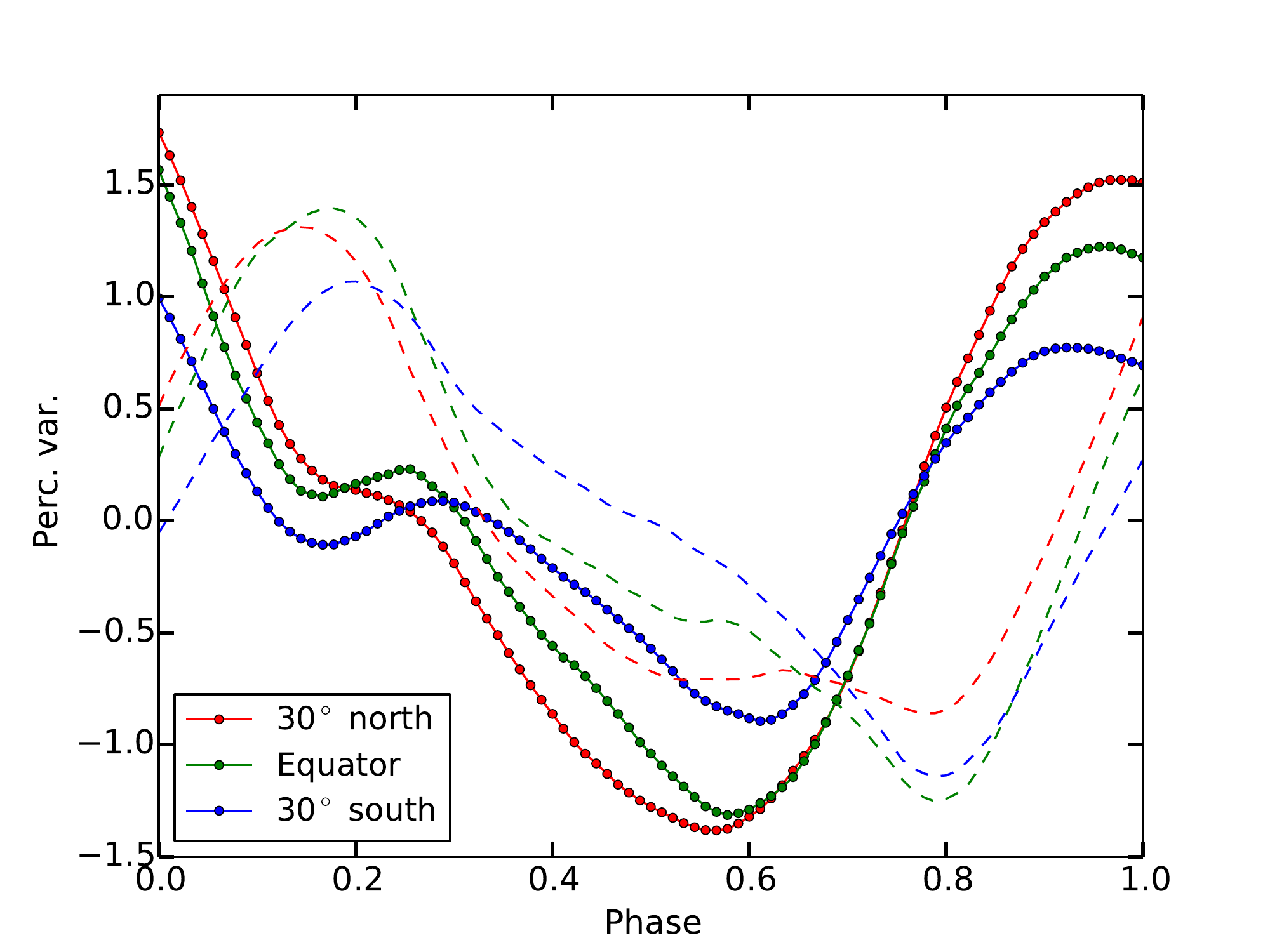}
\caption{Synthetic light curves calculated using the heat flux emanating from the outer boundary of a simulation. Pearled curves are for model A-Pr10-N5 and dashed curves are for the hydrodynamic model H-Pr10-N5. Light-curves are given for three different inclinations: line-of-sight 30$^{\circ}$ north of the equator (red), in the equatorial plane (green), and 30$^{\circ}$ south of the equator (blue). The amplitude is in terms of percentage variations about the mean.}
\label{fig:light_curve}
\end{center}
\end{figure}

Synthetic light curves for the simulation with polar spots (model A-Pr10-N5) and the non-magnetic reference model H-Pr10-N5 are shown in Fig.~\ref{fig:light_curve}. We calculate the light curves for one full rotation following on the instance in time at which the flux is shown in Fig.~\ref{fig:Pr10_dyn_vr_Br}. To generate the light curves the flux at different locations was integrated for the visible hemisphere for different phase angles of rotation. Limb-darkening was incorporated by modulating the visible flux as $f(q)=f_o\,(1-w(1-cos(q)))$ where $w$ is a limb-darkening coefficient (set to a nominal value of 0.3), $f(q)$ is the flux observed by the observer, $f_o$ is the heat flux at a location on the outer boundary of the simulation, $q$ is the angle between the radius vector to a surface point and the line-of-sight. The light curve variations are qualitatively similar for different assumed inclinations of the equatorial plane with respect to the line-of-sight (pearled curves), however, the amplitude increases for equatorial and northern inclinations. We also calculated the light curves for the hydrodynamic simulation H-Pr10-N5 which are shown using dashed curves. These hydrodynamic light curves show a variation of little more than 1\%. The light curve amplitudes for the hydrodynamic model are similar to the amplitude of the southern inclination light curve of the dynamo case where the big dark spot is not visible. Hence, assuming that the model inherently produces light curve modulations of about 1\%, the presence of the big dark spot near the north pole adds an extra modulation of about 0.5\%. Light curve variations of the order of a few percent are consistent with the observation of active cool stars in the {\em Kepler} data set~\citep{basri2013}.

\subsubsection{Importance of density stratification and rotation}
The results discussed above demonstrated that for a fixed density stratification and rotation rate multipolar dynamos (generating weaker field strengths) do not produce dark spots while the ADD dynamos with stronger fields do form dark spots. Using a small parameter study we now explore the sensitivity of dark spot formation to the density stratification in the convection zone and the rotation rate of the spherical shell.

\begin{figure}
\begin{center}
\includegraphics[scale=0.45]{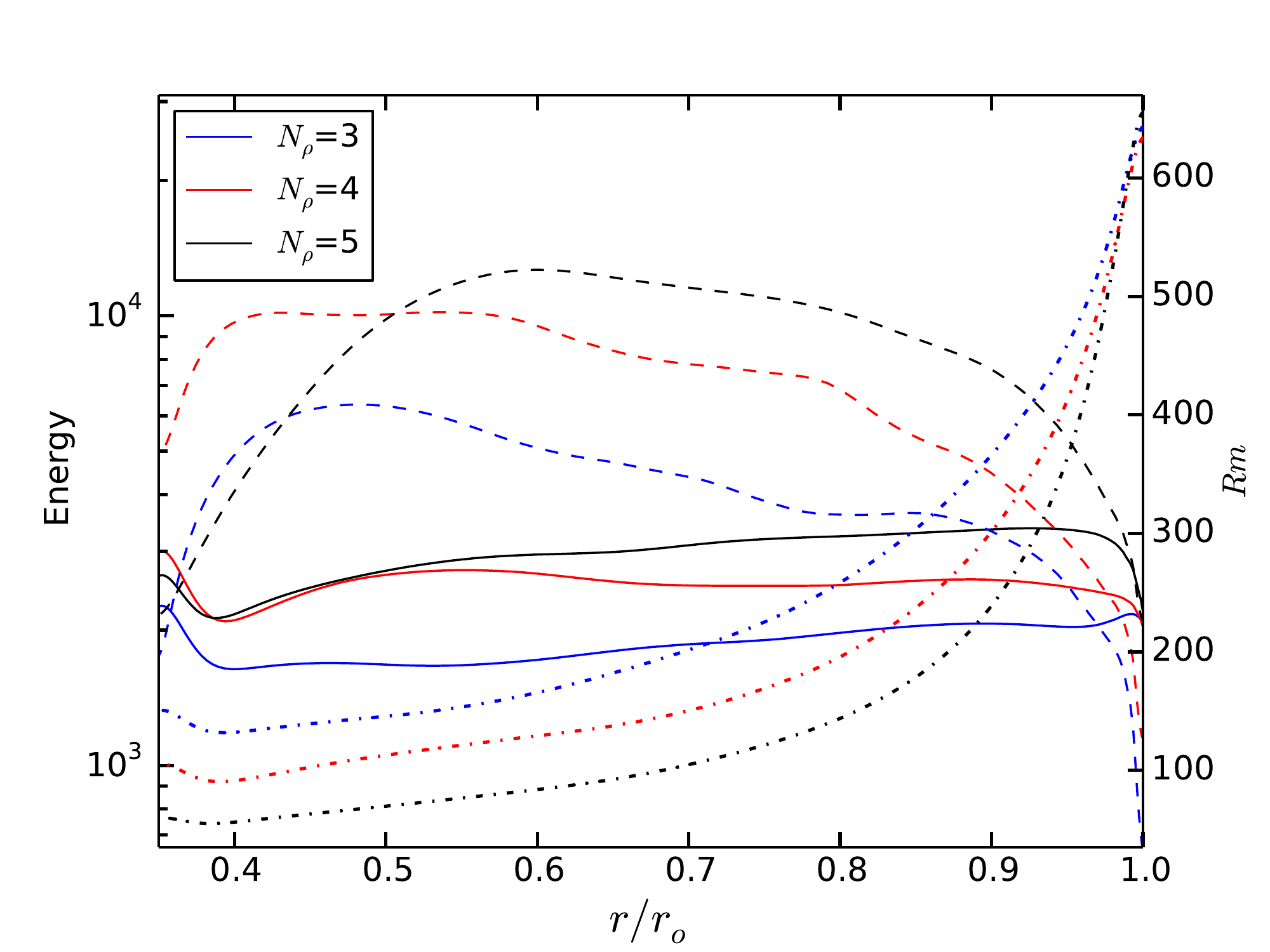}
\caption{Kinetic energy $E_{kin}$ (solid curves), magnetic energy $E_{mag}$ (dashed curves) density on the left axis and magnetic Reynolds number (dotted curves) on the right axis as a function of radius for three models with dipole dominant magnetic fields but different density stratification $N_{\rho}$ in the convection zone. Blue, red, and black  curves are for model A-Pr10-N3, A-Pr10-N4, and A-Pr10-N5, respectively.}
\label{fig:density}
\end{center}
\end{figure}

\begin{figure}
\begin{center}
\includegraphics[scale=0.44]{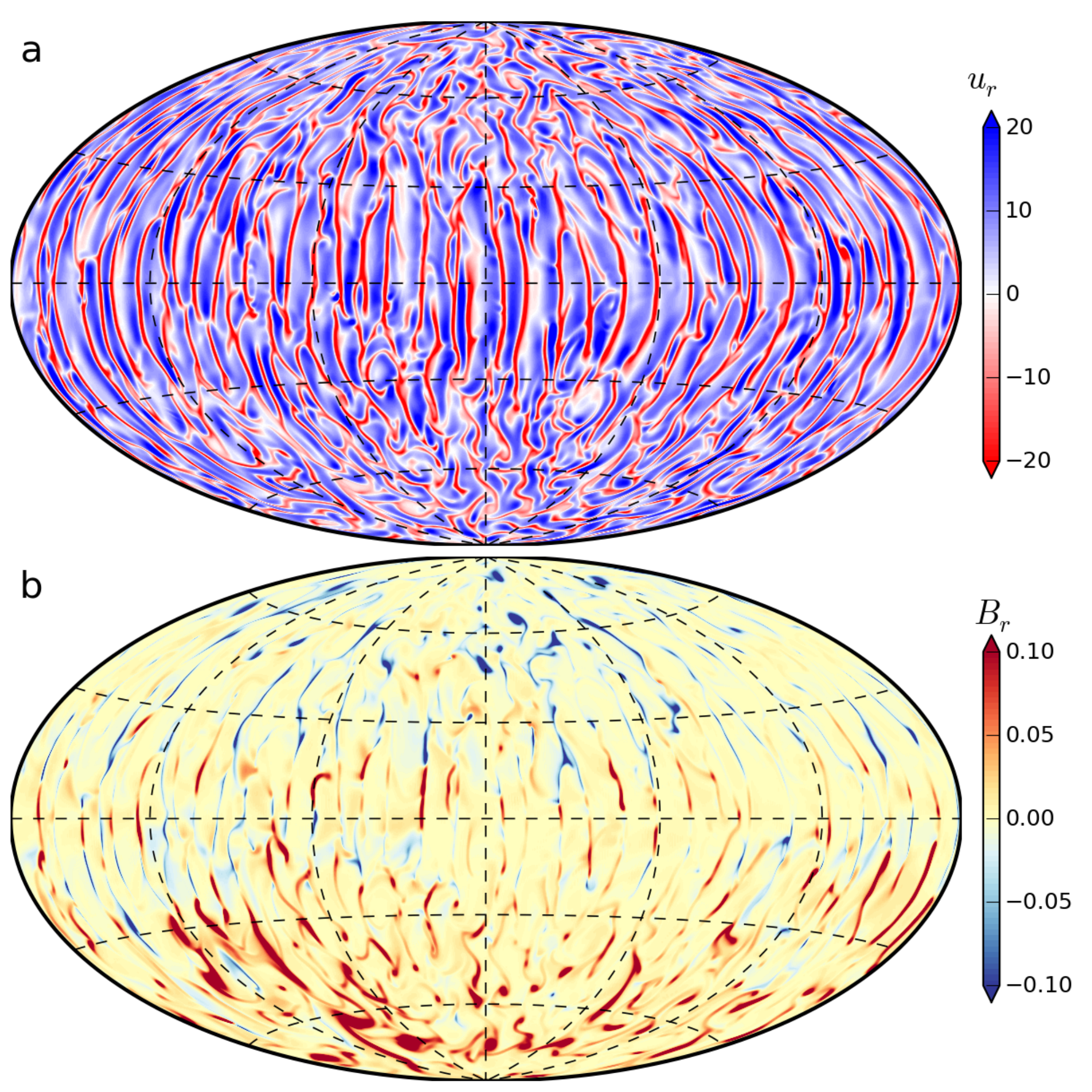}\\
\includegraphics[scale=0.44]{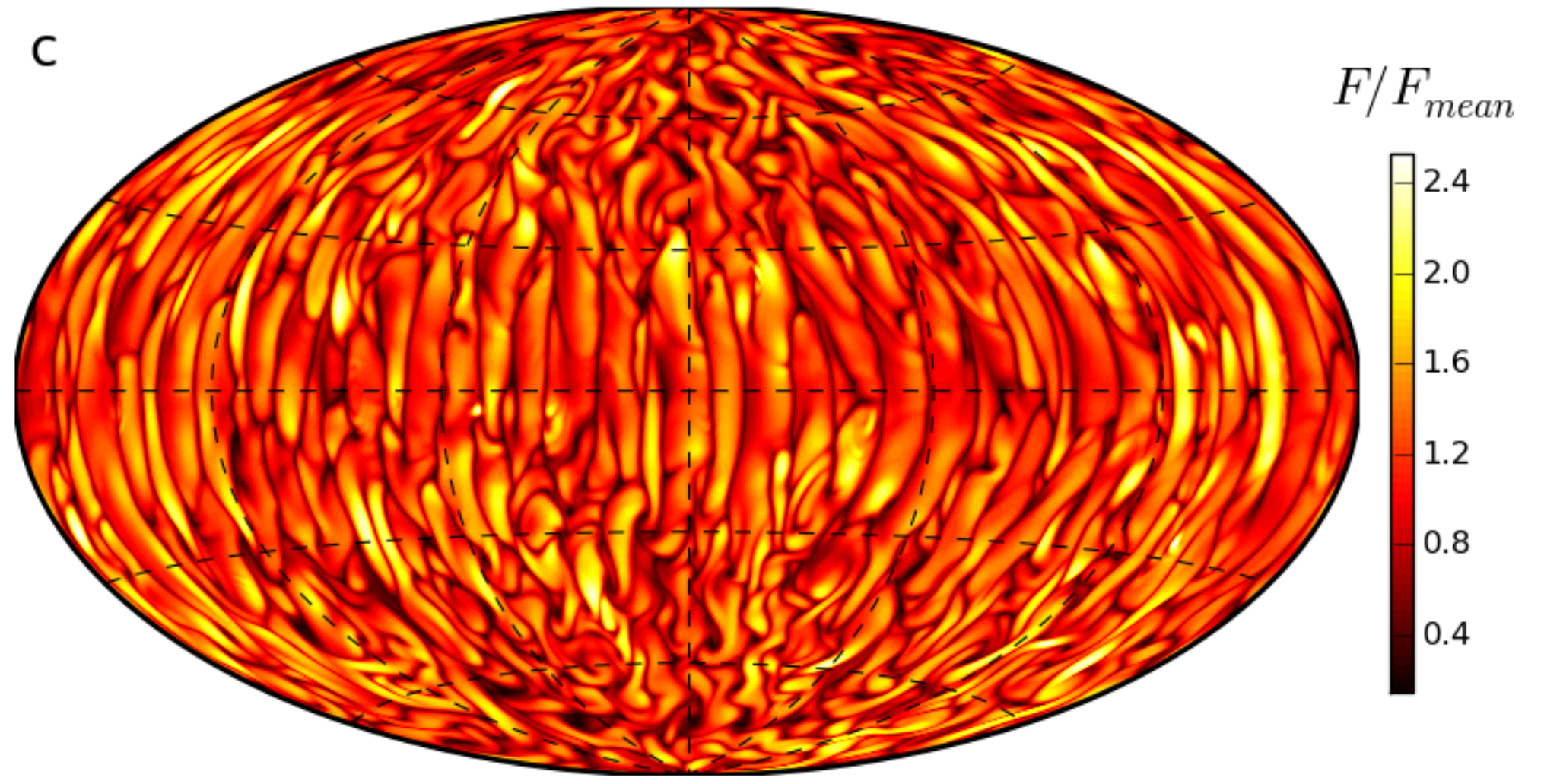}
\caption{Similar to Fig.~\ref{fig:Pr10_dyn_vr_Br} but for model A-Pr10-N3.}
\label{fig:Pr10_dip_vr_Br_N3}
\end{center}
\end{figure}

We simulate two cases with smaller density stratification: model A-Pr10-N4 with $N_{\rho}$=4 (density contrast of 55) and model A-Pr10-N3 with $N_{\rho}$=3 (density contrast of 20). The entropy contrast (or equivalently $Ra$) across the shell is also changed such that the simulations produce an ADD magnetic field (see Tab.~\ref{tab:tab1}). The resulting radial variation of kinetic and magnetic energy is shown in Fig.~\ref{fig:density}. As the density stratification is decreased the  magnetic energy near the outer boundary decreases sharply. Figure \ref{fig:Pr10_dip_vr_Br_N3} displays the same quantities as Fig.~\ref{fig:Pr10_dyn_vr_Br} for model A-Pr10-N3. At a few inter-cellular nodes where the flux is strong enough to locally quench the flow small dark patches are formed, but the surface is devoid of sizeable dark spots. This suggests that inclusion of a large density contrast ($N_{\rho}\geq 5$) in the convection zone is instrumental for generating big and persistent dark spots. However, the mechanism through which density stratification promotes larger flux-concentrations is not yet clear. The proposed "negative effective magnetic pressure instability (NEMPI)"~\citep{rogachevskii2007, brandenburg2011} also highlights the importance of density stratification for generating sizeable magnetic flux concentration. Further analysis is needed to establish a connection between our simulations and the NEMPI mechanism.

\begin{figure}[h]
\begin{center}
\includegraphics[scale=0.44]{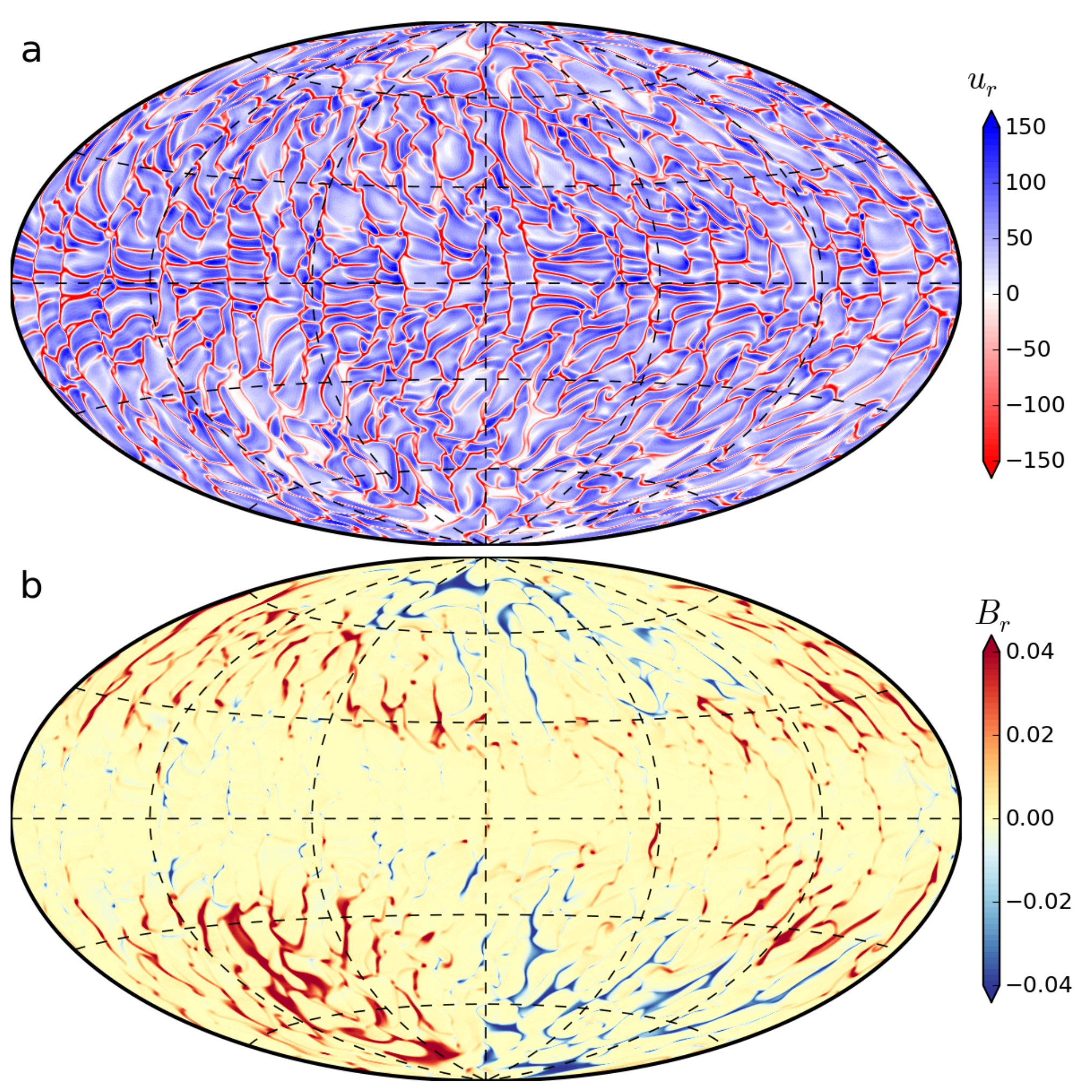}\\
\includegraphics[scale=0.44]{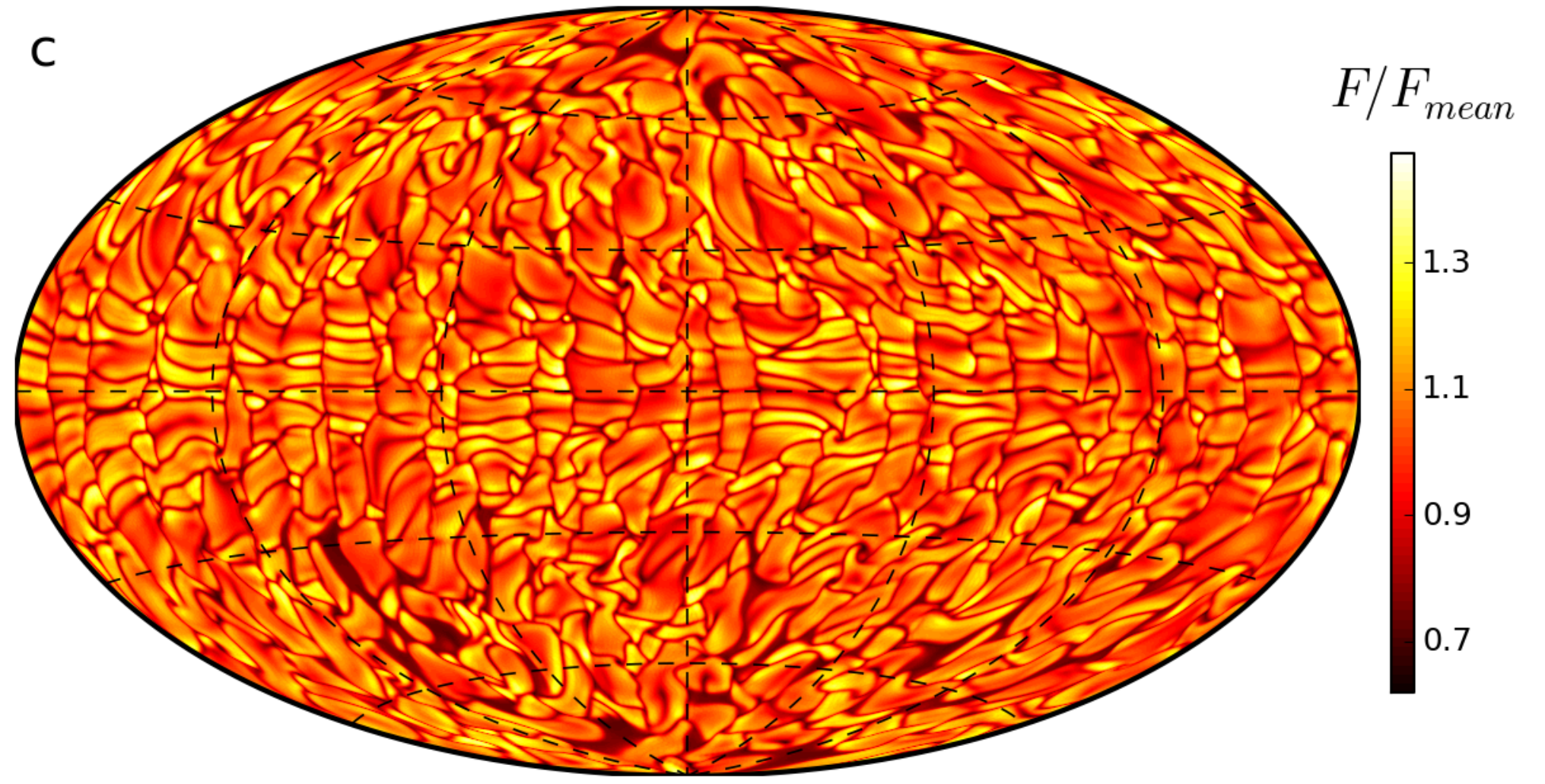}
\caption{Similar to Fig.~\ref{fig:Pr10_dyn_vr_Br} but for model NA-Pr1-N6.}
\label{fig:Pr1_multi_vr_Br}
\end{center}
\end{figure}

Guided by results of the preceding discussion we now increase the density contrast to $N_{\rho}$=6 (density contrast $\approx$400; model NA-Pr1-N6) in the convection zone to explore if a multipolar dynamo can also form dark spots. $Pr$ and $Pm$ in case NA-Pr1-N6 are lowered to avoid long saturation time scales, however, at the expense that a dominantly dipolar magnetic field cannot be sustained (see Tab.~\ref{tab:tab1}). A snapshot, similar to Fig.~\ref{fig:Pr10_dyn_vr_Br}, of dynamo model NA-Pr1-N6 is shown in Fig.~\ref{fig:Pr1_multi_vr_Br}. The magnetic field is dominated by a non-axisymmetric $m$=1 pattern and it peaks in two broad longitudes in both hemispheres. This large scale magnetic field also slowly propagates westwards, similar to the model NA-Pr10-N5. The convection near the outer boundary is dominated by small convection cells, not only at high latitudes as in case A-Pr10-N5, but also near the equator. This is due to the higher $Ro_l$ in this case which varies from 0.05 in the interior to a maximum of about 0.25 in the outer layers.

\begin{figure}
\begin{center}
\includegraphics[scale=0.46]{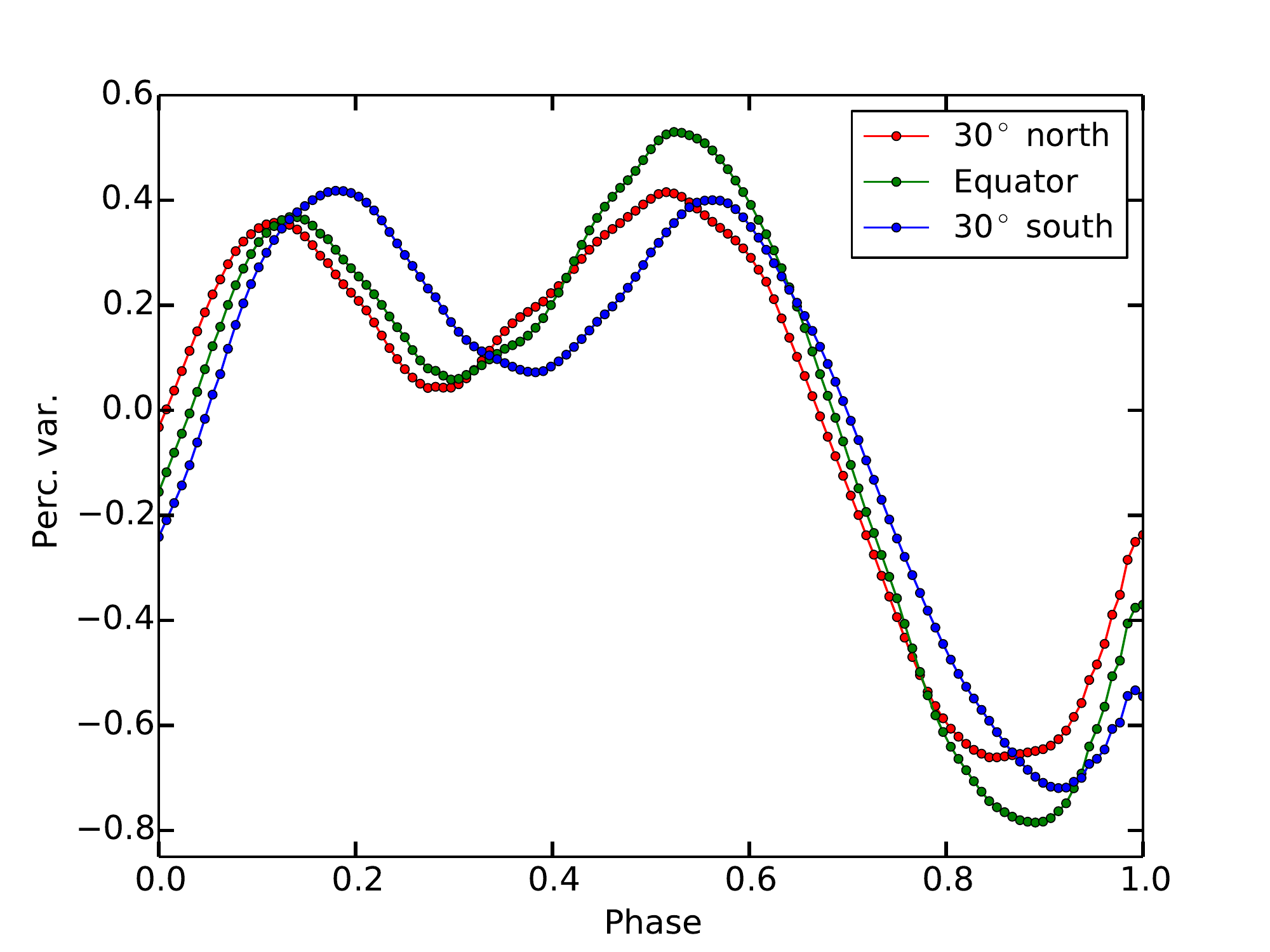}
\caption{Similar to Fig.~\ref{fig:light_curve} but for model NA-Pr1-N6. Light curves are based on the instant shown in Fig.~\ref{fig:Pr1_multi_vr_Br}.}
\label{fig:light_curve2}
\end{center}
\end{figure}

A careful glance at Fig.~\ref{fig:Pr1_multi_vr_Br}({\bf a}) and ({\bf b}) shows that radial velocity is quenched in many localized regions. This gives rise to dark spots that are smaller than the big spot seen in model A-Pr10-N5, but are still quite sizeable, i.e. have the same scale as the  convection cells. The light curve of this case, portrayed in Fig.~\ref{fig:light_curve2}, shows double dips which might be associated with the non-axisymmetric heat flux pattern induced by the $m$=1 magnetic field geometry. Note that the light curve amplitudes in this case are significantly smaller as compared to Fig.~\ref{fig:light_curve} because the convection transports only about 30\% of total heat in this case as compared to about 2/3$^{rd}$ in model A-Pr10-N5. Hence, the dark spots in this case are less contrasting with the regions where the convection is operating normally.  \citet{cole2014} have already reported the existence of similar westward drifting magnetic fields in direct numerical simulations. Such dynamo solutions provide a rather enticing explanation for the "active-longitudes" phenomenon, i.e starspots concentrated in broad preferred longitudes. \citet{cole2014} have  discussed such a connection although their simulation did not generate any dark spots. Our model NA-Pr1-N6 can be considered as a step further in this line of thought since it produces the appropriate magnetic field as well as dark spots.

\begin{figure}
\begin{center}
\includegraphics[scale=0.45]{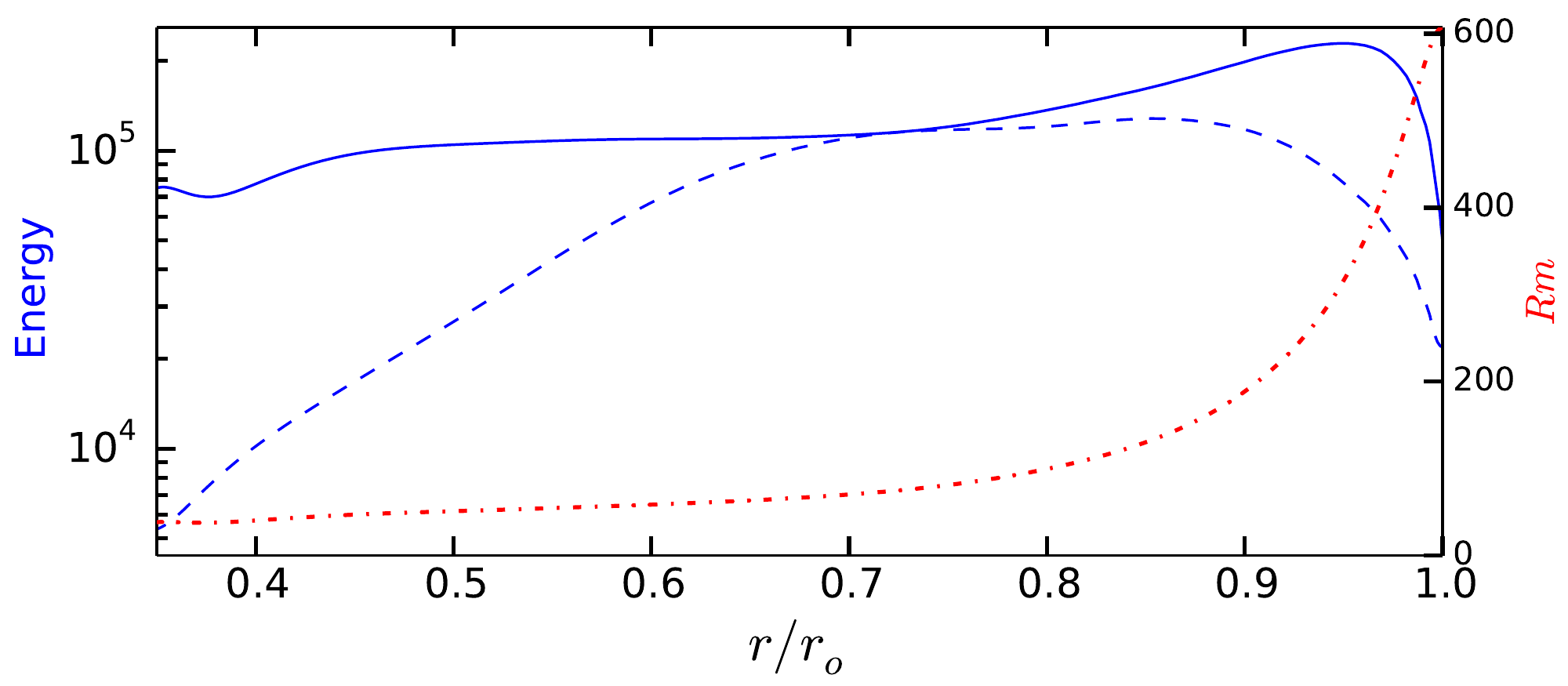}
\caption{Similar to Fig.~\ref{fig:Pr10_dyn_energy} but for model NA-Pr1-N6.}
\label{fig:Pr1_multi_energy}
\end{center}
\end{figure}

The plot Fig.~\ref{fig:Pr1_multi_energy} shows the radial variation of kinetic and magnetic energy density for model NA-Pr1-N6. Model NA-Pr1-N6 is qualitatively similar to model NA-Pr10-N5 (see Fig.~\ref{fig:Pr10_dyn_energy}({\bf a})) in that the kinetic energy dominates in this case as well. However, in outer layers, the dominance of kinetic energy is somewhat smaller: ratio of magnetic to kinetic energy density for model NA-Pr10-N5 and NA-Pr1-N6 are 0.18 and 0.25, respectively, about 40\% larger in the latter. One might speculate that the higher density stratification and less dominance of kinetic energy might explain why model NA-Pr1-N6 generates sizeable dark spots while model NA-Pr10-N5 does not. However, since model NA-Pr10-N5 and NA-Pr1-N6 are also different in other control parameters, narrowing down the main cause for dark spot formation in the latter needs more analysis.

\begin{figure}
\begin{center}
\includegraphics[scale=0.45]{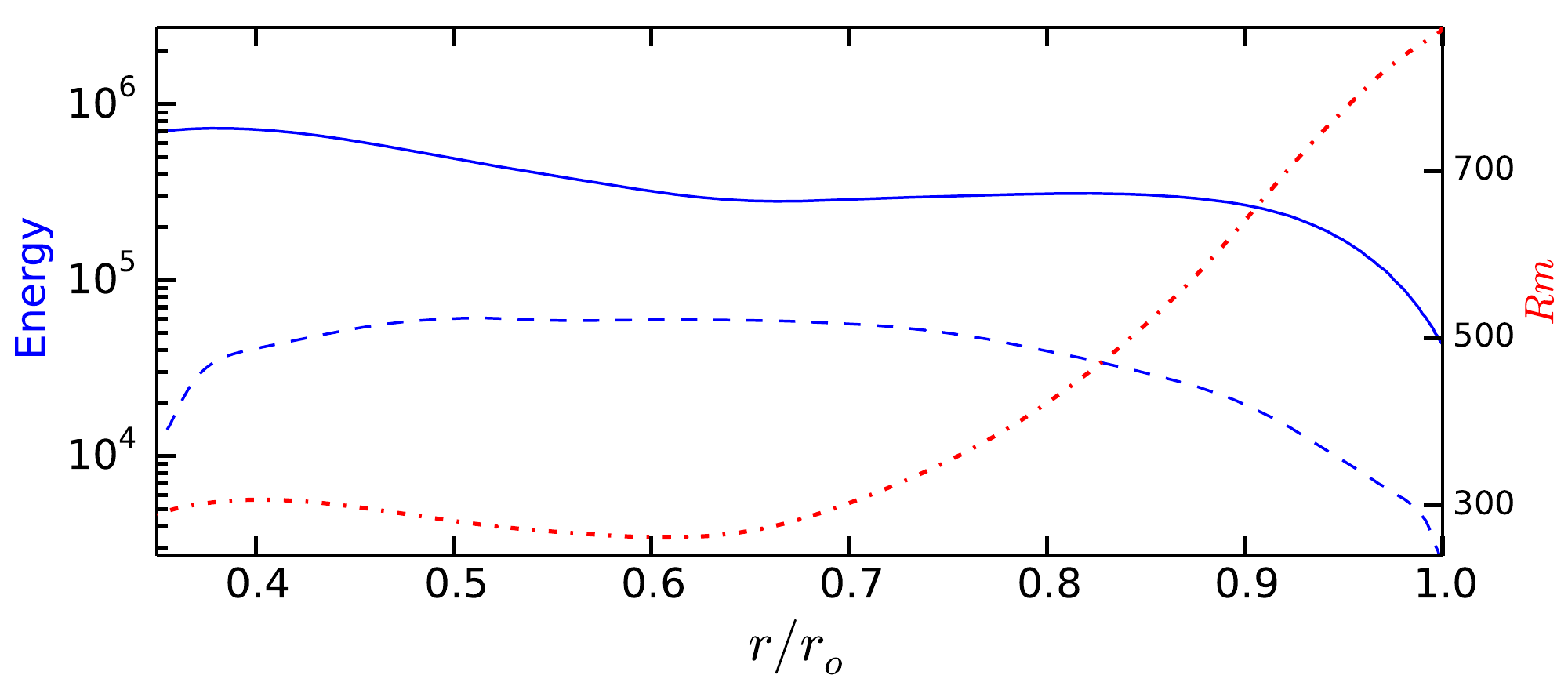}
\caption{Similar to Fig.~\ref{fig:Pr10_dyn_energy} but for model S-Pr1-N5.}
\label{fig:Pr1_small_energy}
\end{center}
\end{figure}

\begin{figure}
\begin{center}
\includegraphics[scale=0.44]{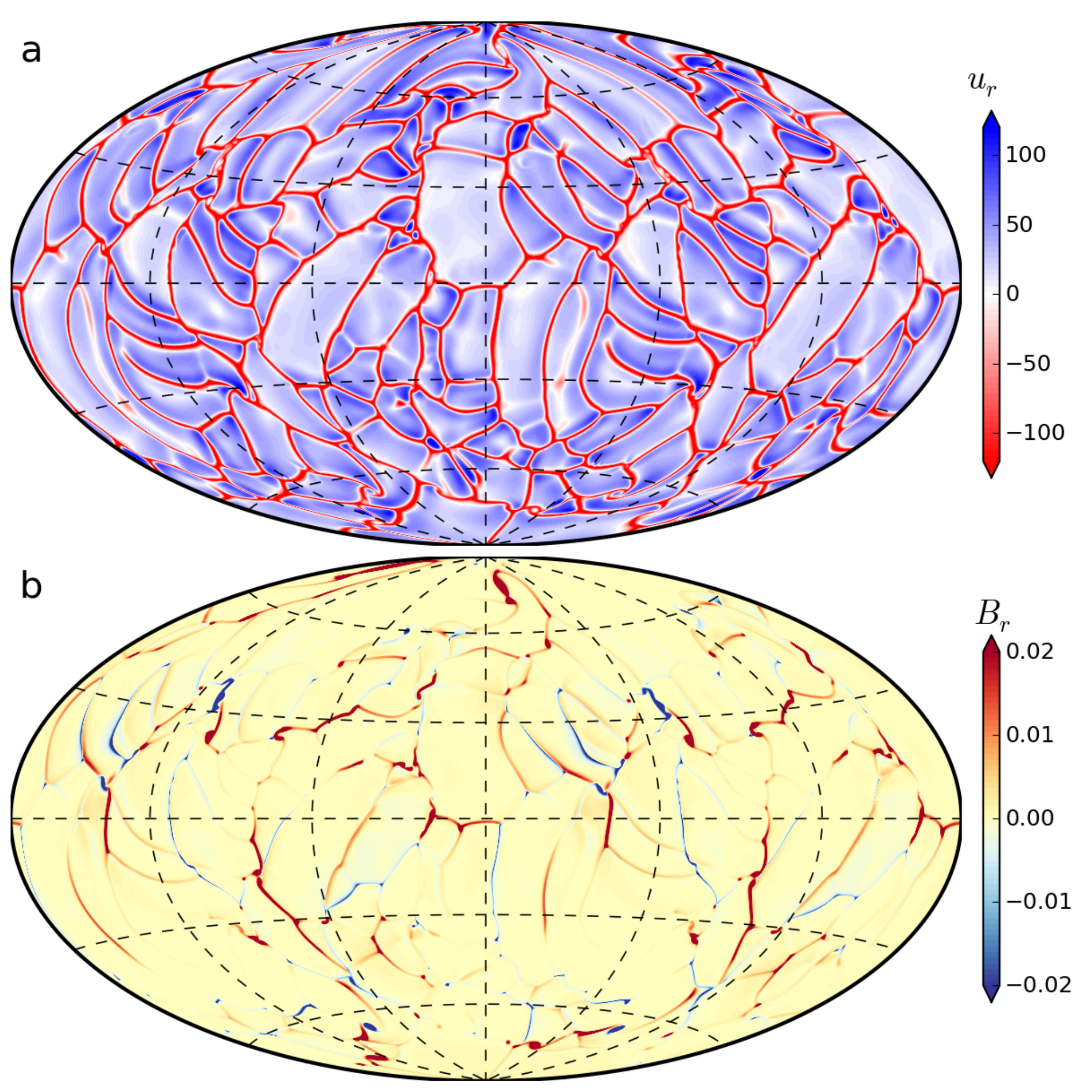}\\
\includegraphics[scale=0.44]{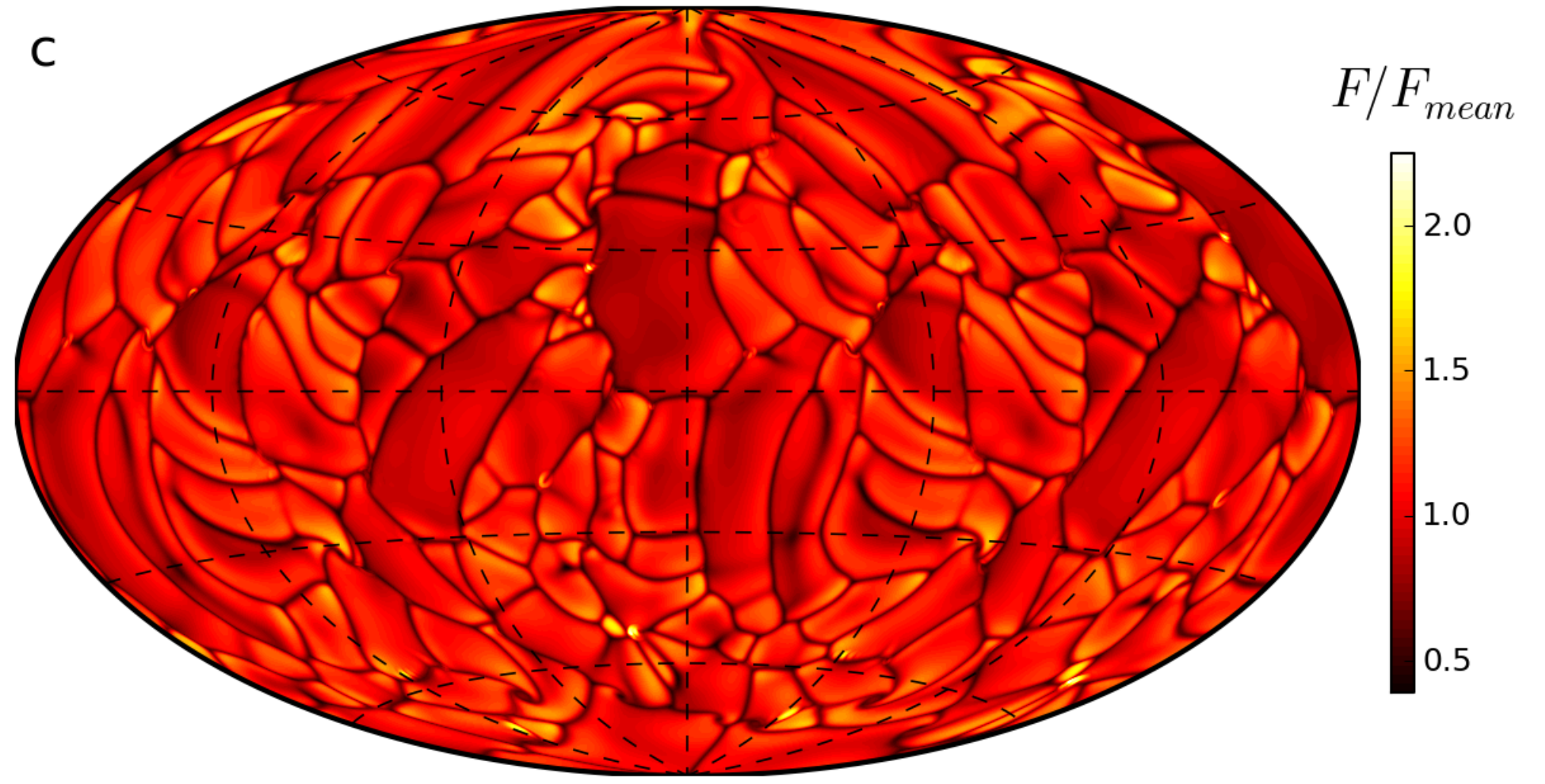}
\caption{Similar to Fig.~\ref{fig:Pr10_dyn_vr_Br} but for model S-Pr1-N5.}
\label{fig:Pr1_small_vr_Br}
\end{center}
\end{figure}

In another set of simulations, we maintained a strong density stratification (density contrast of $\approx$150) but lowered the rotation rate (higher Ekman number) and changed other parameters so that $Ro_l$ is of order one, which puts it into the regime where inertia dominates over rotational forces~\citep{gastine2013a}. No columnar convection exists in this case and the flow is similar to the classical Rayleigh-B{\'e}nard convection (Fig.~\ref{fig:Pr1_small_vr_Br}) and generates anti-solar differential rotation (slower equator and faster poles)~\citep{gastine2014}. The  generated magnetic energy is smaller than the kinetic energy by more than an order of magnitude (Fig.~\ref{fig:Pr1_small_energy}). As shown in Fig.~\ref{fig:Pr1_small_vr_Br}({\bf b}) the magnetic field is very small-scaled and resides mainly in downwellings. No dark spots are observed in this case. Similar observations were also made by \citet{dorch2004} in a dynamo simulation of slowly rotating late-type giant star. However, super-equipartition field strengths were reported in that simulation while we observe fields with rather low strength. 

\section{Summary and outlook}
We have studied the spontaneous formation of dark spots in self-consistent global models of stellar convection and the associated dynamo process. We used fully non-linear anelastic simulations in rotating spherical shells. For rapid rotation, convection is in the form of large axially-aligned helical columns in the interior and more fractured and smaller scaled in the outer layers. A large scale distributed dynamo operates in the bulk of the model. The dynamo generates  magnetic field which interacts with the small scale convective motions in the outer layers where it is swept to the downwellings. This sweeping of flux leads to the formation of localized regions of intense magnetic field. Some of these regions have high enough field strength to locally quench convection, which leads to the formation of dark spots.

Relatively large polar spots were found in a model where the magnetic field has an axial-dipole dominated geometry. In this case moderate-sized magnetic flux concentrations are frequently formed. Occasionally the coalescence of several such flux patches form a big spot. The bigger polar spots were also rather stable and persisted for many tens of rotations. The magnetic flux-bundles associated with the big dark spots were rooted in deeper helical columns. We conjecture that such anchored flux-bundles impose a large length scale and longer time scale on the dark spots associated with them. This scenario is somewhat similar to the one put forth by \citet{kitiashvili2010} who propose that deeper simulation boxes, where larger and slower convection cells coexist with more turbulent ones, are important for having stable and large magnetic features.

The temporal evolution of the dark spots generated on the outer surface of the model was very rich. However, blurring the details of the model and analysing only the large scale modulations hides such temporal behaviour. Such filtering of data produces broad dark regions near the poles which maintain their integrity for hundreds of rotations. This feature is in agreement with  observations~\citep{hussain2002} which indicate that polar dark spots persist for hundreds of stellar rotations. Low-pass filtering of our data also produces low-contrast dark patches all over the surface which do not represent distinct dark spots but seem to reflect large scale heat flux inhomogeneities induced by a mesoscale convective network underneath the surface~\citep{rincon2005,bessolaz2011}. It is, however, not certain if such large scale inhomogeneities (not associated with dark spots) will survive the turbulent photospheric convection which we have not modelled. 

Synthetic light curves of the model with polar spots showed variations with amplitudes reaching around 1.5\%. Comparing light curves at different inclinations suggests that dark spots at high latitudes in the simulation do significantly contribute to the light curve modulations. It should be noted that we do not model the upper stellar layer below the photosphere and omit the top $\approx$5\% of the stellar radius. However, the lateral extent of the large polar spot is much more than the thickness of the unmodelled layer and the strong magnetic field in the dark spot should highly quench convection in this omitted layer as well. Therefore, we expect that the heat flux deficit in the big spot at the surface of our model represents that at the photosphere.

Based on a small parameter study, the following ingredients seems to be crucial for sizeable dark spot formation: 1) rotation-dominated convection, which favours the generation of a large scale dynamo; 2) high density stratification (five or more density scale heights) in the convection zone, producing cellular convection along with deeper large-scale columns; 3) dynamos with axial-dipole dominated field which inherently produces strong and {\em stable} magnetic field, allowing dark spots to grow to bigger sizes.

The result in this study are based on a non-dimensional formulation of the anelastic-MHD equations. In a generic sense they can be applied to different types of stars as long as rotation plays an important role. Examples include young pre-main sequence stars, evolved giants or low-mass stars with small radiative cores. However, it can also be instructive to express our results in an exemplary way in physical units. Here we scale the model with the big polar spot (A-Pr10-N5) to a pre-main sequence star with 0.7 solar masses and an effective surface temperature of 4200 K (spectral type K5). For the internal structure of such object we use results from a stellar evolution model \citep{granzer2000}.  We determine physical values at a radius $R_o$ somewhat below the photosphere (at about 0.96 R) in the stellar model where the density has dropped by a factor of 150 below its value at 0.34R. Using the physical values from the interior model and matching the heat flux at $R_o$ to the outer boundary heat flux of the polar spot case provides $\nu$ = 4.3 $\times$10$^8$ m$^2$~s$^{-1}$, $\lambda$ = $\kappa$ = 4.3$\times$10$^7$ m$^2$~s$^{-1}$ and $\Omega$ = 5.6$\times$10$^{-6}$ s$^{-1}$ for the rotation rate. As usual for such simulations, the diffusivities are much higher than molecular values and must be understood as effective turbulent diffusivities. The rotation period is 13 days. This is not by itself a very fast rotation rate, however, from a force balance argument a system qualifies as rapidly rotating when the Rossby number is much less than one, which is the case in our model. With these physical inputs, the variation range for the magnetic field in Fig.~\ref{fig:Pr10_dyn_vr_Br}({\bf d}) is about $\pm 15$kG. The unsigned surface radial field averages to about 1.4 kG which is a typical magnetic field strength found at the surface of rapidly rotating (Rossby number $<0.1$) stars~\citep{reiners2009, vidotto2014}. Assuming that the heat flux variations (at least those associated with big spots) produced on the outer boundary of our simulation propagate to the stellar photosphere, we calculate from the Stefan-Boltzmann law an effective surface temperature that varies in the range 3000 K < $T_{eff}$ < 5000 K over the stellar surface. Considering only the large scale component of the heat flux variations (as shown in Fig.~\ref{fig:Pr10_dip_flux_trunc}({\bf b})) narrows this range to 3700 K < $T_{eff}$ < 4400 K. The temperature anomaly of the order 500 K is comparable to what has been inferred observationally for stars compatible with the spectral type chosen here~\citep{berdyugina2005}.

In stellar convection zones most of the heat is transported by the vigorous convective motions. Our models are not in such a turbulent state and a substantial amount of heat in the models is carried by diffusive processes. The magnetic quenching of convection is strong in our simulations, producing large variations in the heat flux carried by convection. However, the associated modulations in the {\em total} heat flux are somewhat moderate, reaching about 60\% below surface average in the darkest spots. Hence, the range of temperature modulations mentioned above (based on heat flux variations) should be considered as lower estimates.

We have omitted many ingredients in our models which can affect the results. The presence of a sub-adiabatic radiative core might introduce additional interesting dynamics~\citep{brown2010, ghizaru2010, masada2013}. The inclusion of a sub-adiabatic coronal region might promote local bipolar structures~\citep{warnecke2013}. It is not clear how a photospheric small-scale dynamo~\citep{vogler2007} might affect the dark spots produced by an interior large-scale dynamo. A multitude of interesting phenomenon, e.g. formation of plage, related to the modelling of opacity might change the nature of the light-curves we calculated. Furthermore, following up on the strategies outlined here using fully compressible approaches \citep[e.g.][]{kapyla2012} would help to further confirm the startspot formation mechanism presented here.

The choice of a relatively large (magnetic) Prandtl number in the polar spot case can be questioned. The need for high values of $Pr$ and $Pm$ in order to stabilize ADD magnetic field is probably a consequence of using high values of the Ekman number in density-stratified stellar dynamo simulations (due to technical constraints). Similar to the planetary dynamo simulations, it should also be possible to find ADD dynamos in stellar dynamo simulations at low $Pr$ and $Pm$ once the Ekman number is low enough.

Our model is simplified in several respects and cannot address many of the details of the formation and the properties of starspots. Nonetheless, it is to our knowledge the first global model that generates rather big polar spots in a completely self-consistent way. It points at an interesting alternative to the flux-tube model of spot formation by a distributed dynamo mechanism.

\begin{acknowledgements} 
We thank the referee for a very constructive review and Manfred Sch{\"u}ssler, Robert Cameron, J\"orn Warnecke for commenting on the manuscript. We also thank Jayant Joshi, Julien Morin, and Laur\`ene Jouve for interesting discussions. Funding from the DFG through SFB 963/A17 and SPP 1488 is acknowledged. Computations were performed on RZG, HLRN (project ``nip00031"), and GWDG. Diligent efforts by Tilman Dannert (RZG) who implemented MPI and then hybrid OpenMP+MPI parallelisation in MagIC are greatly appreciated. 
\end{acknowledgements}


\end{document}